\documentclass[iop,usenatbib]{emulateapj}

\usepackage[english]{babel}
\usepackage{graphicx}
\usepackage{fleqn}
\usepackage{amssymb}
\usepackage{amsmath}
\usepackage{booktabs}
\usepackage{threeparttable}
\usepackage{enumerate}
\usepackage[breaklinks,colorlinks,citecolor=blue]{hyperref}
\usepackage{comment}

\usepackage{txfonts}

\renewcommand{\S}{Section}
\newcommand{\F}{Fig.}

\newcommand{\ve}[1]{\mathbf{#1}}
\newcommand{\unit}[1]{\hat{\mathbf{#1}}}

\newcommand{\Ec}{\mathcal{E}}

\newcommand{\msun}{\mathrm{M}_\odot}

\newcommand{\au}{\,\textsc{au}}

\newcommand{\oin}{\mathrm{in}}
\newcommand{\oout}{\mathrm{out}}

\newcommand{\adpar}{\mathcal{R}}

\newcommand{\bJ}{\mathbf{J}}

\newcommand{\rr}{\mathrm{r}}
\newcommand{\rv}{\mathrm{v}}
\newcommand{\ri}{\mathrm{i}}

\newcommand{\rc}{\mathrm{c}}

\newcommand{\Tcvrr}{t_\mathrm{coh}^\mathrm{v}}

\usepackage{etoolbox}
\newtoggle{ApJFigs}
\togglefalse{ApJFigs}

\begin{document}

\title{The impact of vector resonant relaxation on the evolution of binaries near a massive black hole: implications for gravitational wave sources}
\author{Adrian S. Hamers$^{1}$, Ben Bar-Or$^{1}$, Cristobal Petrovich$^{2,3}$, Fabio Antonini$^{4}$}
\affil{$^{1}$Institute for Advanced Study, School of Natural Sciences, Einstein Drive, Princeton, NJ 08540, USA \\
$^{2}$Canadian Institute for Theoretical Astrophysics, University of Toronto, 60 St George Street, ON M5S 3H8, Canada \\
$^{3}$Centre for Planetary Sciences, Department of Physical \& Environmental Sciences, University of Toronto at Scarborough, Toronto, Ontario M1C 1A4, Canada 
$^{4}$Astrophysics Research Group, Faculty of Engineering and Physical Sciences, University of Surrey, Guildford, Surrey, GU2 7XH, United Kingdom}
\email{hamers@ias.edu}
\email{benbaror@ias.edu}
\email{cpetrovi@cita.utoronto.ca}
\email{f.antonini@surrey.ac.uk}

\begin{abstract}    Binaries within the sphere of influence of a massive black hole (MBH) in galactic nuclei are
  susceptible to the Lidov-Kozai (LK) mechanism, which can drive orbits to
  high eccentricities and trigger strong interactions within the binary such as the emission of
  gravitational waves (GWs), and mergers of compact objects. These
  events are potential sources for GW detectors such as Advanced LIGO and VIRGO\@. The
  LK mechanism is only effective if the binary is highly inclined with respect
  to its orbit around the MBH (within a few degrees of $90^\circ$), implying low rates. However,
  close to an MBH, torques from the stellar cluster give rise to the process of
  vector resonant relaxation (VRR). VRR can bring a low-inclination binary into
  an `active' LK regime in which high eccentricities and strong
  interactions are triggered in the binary. Here, we study the coupled LK-VRR dynamics, with
  implications for LIGO and VIRGO GW sources. We carry out Monte Carlo
   simulations and find that the merger fraction enhancement due to LK-VRR dynamics is
  up to a factor of $\sim 10$ for the lower end of assumed MBH masses
  ($M_\bullet = 10^4\,\msun$), and decreases sharply with increasing
  $M_\bullet$. We find that, even in our most optimistic scenario, the baseline 
  BH-BH merger rate is small, and the enhancement by LK-VRR coupling is not
  large enough to increase the rate to well above the LIGO/VIRGO lower limit, $12\,\mathrm{Gpc^{-3}\,yr^{-1}}$.
  For the Galactic Center, the LK-VRR-enhanced rate is $\sim 100$ times lower than the LIGO/VIRGO limit, and for 
  $M_\bullet = 10^4\,\msun$, the rate barely reaches $12\,\mathrm{Gpc^{-3}\,yr^{-1}}$.
\end{abstract}


\section{Introduction}
\label{sect:introduction}
The recent direct detections of gravitational waves from merging black holes
(BHs) and neutron stars (NSs)
\citep{2016PhRvL.116x1103A,2016PhRvL.116f1102A,2017PhRvL.118v1101A,2017ApJ...851L..35A,2017PhRvL.119n1101A,2017ApJ...848L..12A}
have ushered in the era of gravitational wave (GW) astronomy. One of the key
astrophysical questions motivated by these observations relates to the
formation of BH and NS binaries: the progenitor stars were much larger, so how could they evolve to become compact objects and be
driven close enough to each other to coalesce? A number of formation channels
have been proposed and investigated, including isolated binary evolution, or
other processes in non-isolated binary systems. In the former case, a high-mass
stellar binary is driven by common-envelope evolution or non-conservative mass
transfer to a short period, i.e., sufficiently short for the binary to merge
within a Hubble time due to energy loss by GW radiation (e.g.,
\citealt{1973NInfo..27...70T,1993MNRAS.260..675T,2003MNRAS.342.1169V,2007PhR...442...75K,
2012ApJ...759...52D,2013ApJ...779...72D,2014ApJ...789..120B,2016Natur.534..512B,2017arXiv170607053B}). Alternatively,
rapidly rotating stars in tight binaries can be chemically mixed, preventing
their merger early in the evolution, and resulting in BH mergers
\citep{2016MNRAS.458.2634M,2016AA...588A..50M,2016MNRAS.460.3545D}.

In most `isolated' binary channels, the orbit at the time of merging is
expected to be circular. However, channels involving dynamical environments
have been proposed, some of which predict the BH binary in the LIGO band to have a residual
and non-negligible eccentricity when reaching the LIGO band, thereby distinguishing themselves
from other formation scenarios. With current techniques based on circular
 GW strain templates, the maximum eccentricity for a source in the LIGO band to be detectable is 
 $\approx 0.1$ \citep{2010PhRvD..81b4007B,2013PhRvD..87d3004E,2013PhRvD..87l7501H,
 2018arXiv180605350L}.
 Dynamical formation channels include isolated triple systems
\citep{2017ApJ...836...39S,2017ApJ...841...77A,2017ApJ...846L..11L,2018arXiv180503202L},
mergers arising from interactions in dense clusters (e.g.,
\citealt{1993Natur.364..423S,2000ApJ...528L..17P,2006ApJ...637..937O,2014MNRAS.441.3703Z,2015PhRvL.115e1101R,
2016PhRvD..93h4029R,2016MNRAS.463.2443K,2016MNRAS.459.3432M,
2017ApJ...840L..14S,2017arXiv171206186S,2017arXiv170603776S,2018PhRvD..97j3014S,2018PhRvL.120o1101R}),
and binaries in stellar systems dominated by a central massive BH (MBH\@;
\citealt{2012ApJ...757...27A,2014ApJ...781...45A,2015ApJ...799..118P,2016MNRAS.460.3494S,
2016ApJ...831..187A,2017ApJ...846..146P,2017arXiv170905567A,2018MNRAS.477.4423A,2018ApJ...856..140H,Randall+2018a,
Randall+2018b,2018ApJ...860....5G}). The eccentricity in the LIGO band predicted by these channels
ranges widely, between typically $\sim 10^{-6}$ for cluster models (e.g., \citealt{2016PhRvD..93h4029R}),
$\sim 10^{-3}$ for triple star models (e.g., \citealt{2017ApJ...841...77A}), and $\sim 0.1$ for galactic nuclei (e.g., \citealt{2018ApJ...860....5G}).
Note that some of these models also predict very high eccentricities, i.e., much higher than 0.1 (e.g., due to binary-single
interactions in globular clusters, \citealt{2018PhRvD..97j3014S,2018PhRvL.120o1101R})
, which would likely be missed by current detectors.

In the case of binaries dominated by a central massive BH, the torque of the MBH can accelerate the merging process
through Lidov-Kozai oscillations
(\citealt{1962PSS....9..719L,1962AJ.....67..591K}; see
\citealt{2016ARAA..54..441N} for a review). The LK dynamics of binaries around
MBHs are complicated by the existence of a large range of physical processes
taking place at comparable time-scales such as mass precession, relativistic precession, and 
processes related to stellar relaxation. These processes depend on the binary
separation, the distance of the binary to the MBH, the MBH mass, and the
properties of the stellar cluster around the MBH (see, e.g., \citealt{2012ApJ...757...27A} for an overview).

In most previous studies, the binary+MBH system was treated as an isolated
three-body system, and other processes related to relaxation with other stars in the
cluster around the MBH were not taken into account directly. Recently,
\citet{2016ApJ...828...77V} and \citet{2017ApJ...846..146P} took the first next
step by self-consistently treating effects associated with stellar relaxation
on the evolution of the binary+MBH system. 

In particular, \citet{2017ApJ...846..146P} showed that, if the stellar cluster around the
MBH is nonspherical, then the associated torques on the binary system around the 
MBH can give rise to chaotic secular dynamics, resulting in enhanced eccentricity excitation
in the binary and increased rates of BH mergers, assuming that the stellar cluster is
sufficiently nonspherical. These dynamics are analogous to secular chaotic
dynamics in hierarchical quadruple systems, i.e., with an additional fourth
body instead of a nonspherical background potential
\citep{2017MNRAS.470.1657H,2018MNRAS.474.3547G}.

In this paper, we focus on a physical process that applies even in the
absence of significant nonsphericity of the stellar cluster around the MBH.
Close to an MBH (within its sphere of influence), the motion is nearly Keplerian, and
encounters between stars are therefore correlated. These correlated encounters give
rise to torques that can change both the direction and magnitude of
the angular momentum of the orbit of the binary around the MBH (i.e., the
`outer orbit'), in a process called resonant relaxation (RR\@;
\citealt{1996NewA....1..149R,2006ApJ...645.1152H,2009ApJ...698..641E,2009ApJ...702..884P,2011PhRvD..84d4024M,
2011MNRAS.412..187K,2013ApJ...763L..10A,2014CQGra..31x4003B,2014MNRAS.443..355H,
2015MNRAS.448.3265K,2016MNRAS.458.4129S,2016MNRAS.458.4143S,2016ApJ...820..129B,Fouvry+2017a,2018arXiv180208890B}). The
effect of RR on the magnitude of the angular momentum vector is known as scalar
RR (SRR), whereas the change of the direction of the angular momentum vector is
associated with vector RR (VRR). Both SRR and VRR can be important for the
secular binary+MBH evolution. First, a change of the direction of the outer
orbital angular momentum vector can induce a mutual inclination between the
inner and outer orbits, even if they were initially (close to)
coplanar. Subsequently, a mutual inclination can drive high-eccentricity LK
oscillations in the inner binary. Second, the torque of the outer binary on the
inner binary increases with the outer orbit eccentricity $e_\oout$, increasing
the efficiency (i.e., decreasing the time-scale, see equation~\ref{eq:t_LK}, and
increasing the strength, see equation~\ref{eq:eps_oct}) of the LK mechanism.

The effects of RR on the binary+MBH system were
investigated recently by \citet{2016ApJ...828...77V} using a hybrid $N$-body
technique, in which a small stellar cluster (307 to 4400 stars) around a
massive object ($10^3$ to $10^4$ $\msun$) was integrated, and its output was
used to perturb the binary+MBH system in separate 3-body integrations. Although
innovative with this technique and the first to incorporate RR
with the binary+MBH evolution, the simulations of \citet{2016ApJ...828...77V}
were limited in terms of low particle numbers and low central MBH masses due to
computational limitations. Also, because direct $N$-body techniques were used,
not much insight could be gained into the fundamental dynamics. Here, we adopt
a less computationally expensive approach in which the binary+MBH system is
integrated with a secular code, and the effects of VRR are
incorporated by adopting a simplified model. This approach allows us to more
efficiently disentangle the different physical processes; in particular, it
allows us to investigate the importance of details of VRR-related dynamics 
and the impact on the long-term secular evolution of the binary+MBH system. 
Furthermore, our approach is not limited in terms of MBH mass, and its speed allows 
for the calculation of a large number of systems to explore larger regions of the parameter space.
We note that our focus is on VRR; SRR, although potentially important for the long-term
LK evolution of binaries around an MBH, is beyond the scope of this paper.

The plan of the paper is as follows. In \S\,\ref{sect:time_scales}, we give an
overview of the time-scales of interest for binaries around MBHs, and identify
potential regions of interest for enhanced rates of BH and NS mergers induced
by LK-VRR coupling. We describe a model to take into account the effect of VRR 
on the outer orbit in \S\,\ref{sect:meth}. In \S\,\ref{sect:dyn}, we explore how VRR can 
affect the LK dynamics of binaries around MBHs. We carry out detailed Monte Carlo
calculations including other effects such as post-Newtonian (PN) corrections
and binary evaporation, and evaluate the enhancement of merger rates due to
LK-VRR coupling in \S\,\ref{sect:pop_syn}. We discuss our results in
\S\,\ref{sect:discussion}, and conclude in \S\,\ref{sect:conclusions}.

\section{Time-scales}
\label{sect:time_scales}
Galactic nuclei with a central MBH are complex dynamical environments, and
various physical processes are relevant to the evolution of binaries within
these systems. First, we give a brief overview of the important time-scales
(similar time-scale calculations can be found in, e.g., Section 3 of
\citealt{2012ApJ...757...27A}; for a general overview of galactic
nuclei dynamics, we refer to \citealt{2013degn.book.....M}). We then compare the various
time-scales for different cluster and binary properties, and explore the
regimes in parameter space in which LK-VRR coupling could potentially be
important for enhancing binary merger rates (a more detailed investigation is
carried out in Sections~\ref{sect:dyn} and~\ref{sect:pop_syn}).

An overview of the notation used in this paper is given in Table~\ref{table:notation}. 
Throughout, we assume, for simplicity, a spherically symmetric cluster around the
MBH with a typical stellar mass $m_\star \equiv \langle m^2\rangle^{1/2}$ 
(i.e., $m_\star$ is the square root of the average of the squared stellar mass in the cluster), and
a simple lower-law density distribution, i.e., $\rho_\star(r) \propto r^{-\gamma}$. For
a Salpeter mass function, $\mathrm{d}N/\mathrm{d}m\propto m^{-2.35}$, \citep{1955ApJ...121..161S},
and assuming lower and upper mass limits of $0.08$ and $80$ $\msun$, respectively, 
$\langle m^2\rangle^{1/2} \approx 1.1\,\msun$. For a top-heavy mass function,
$\mathrm{d}N/\mathrm{d}m \propto m^{-1.7}$, and with the same mass ranges, 
$\langle m^2\rangle^{1/2} \approx 5.3 \, \msun$. For simplicity, we consider only the case of
 a Salpeter mass function for the stellar cluster in our simulations, and set
 $m_\star=1\,\msun$. 

\begin{table}
\begin{tabular}{lp{5.0cm}}
  \toprule
  Symbol & Description \\
  \midrule
  $G$						& Gravitational constant. \\
  $c$						& Speed of light. \\
  \midrule
  \multicolumn{2}{l}{Stellar cluster} \\
  $r$						& Distance from the MBH\@. \\
  $M_\bullet$				& Mass of the MBH\@. \\
  $m_\star$					& Average background stellar mass. \\
  $a$						& Semimajor axis of an orbit around the MBH\@. \\
  $e$						& Eccentricity of an orbit around the MBH\@. \\
  $P$						& Orbital period of an orbit around the MBH\@. \\
  $n_\star(r)$				& Stellar number density, assumed to be $n_\star \propto r^{-\gamma}$. \\
  $\rho_\star(r)$				& Stellar mass density, $\rho_\star(r) = m_\star n_\star(r)$. \\
  $\sigma_\star(r)$			& One-dimensional stellar velocity dispersion. \\
  $N_\star(a)$				& Number of stars within radius $r=a$ from the MBH\@. \\
  $M_\star(a)$				& Mass of stars within radius $r=a$ from the MBH\@. \\
  $\ve{J}$					& Angular-momentum vector of an orbit around the MBH\@; $J=\sqrt{GM_\bullet a(1-e^2)}$ if the stellar potential is neglected. \\
  $J_\mathrm{c}$				& Angular momentum of a circular orbit around the MBH\@. \\
  $\gamma$				& Density slope ($\rho_\star \propto r^{-\gamma}$). \\
  \midrule
  \multicolumn{2}{l}{Binary+MBH triple system} \\
  $M_1$					& Inner orbit primary mass. \\
  $M_2$					& Inner orbit secondary mass. \\
  $a_\oin$					& Inner orbit semimajor axis. \\
  $a_\oout$					& Outer orbit semimajor axis. \\
  $P_\oin$					& Inner orbital period. \\
  $P_\oout$					& Outer orbital period. \\
  $e_\oin$					& Inner orbit eccentricity. \\
  $e_\oout$					& Outer orbit eccentricity. \\
  $\ve{e}_\oin$				& Inner orbit eccentricity vector. \\
  $\ve{e}_\oout$				& Outer orbit eccentricity vector. \\
  $\ve{j}_\oin$			& Inner orbit normalized angular-momentum vector; its norm is $j_\oin = \sqrt{1-e_\oin^2}$. \\
  $\ve{j}_\oout$			& Outer orbit normalized angular-momentum vector. \\
  $i_\mathrm{rel}$			& Relative inclination between the inner and outer orbits ($\cos i_\mathrm{rel} = \unit{j}_\oin \cdot \unit{j}_\oout$). \\
  $\omega_\oin$				& Inner orbit argument of periapsis. \\
  $\omega_\oout$			& Outer orbit argument of periapsis. \\
  \bottomrule
\end{tabular}
\caption{Description of symbols used throughout this paper. }
\label{table:notation}
\end{table}

\subsection{Mass precession}
\label{sect:time_scales:MP}
Distributed mass enclosed within an orbit around the MBH causes the orbit to
precess in its plane, a process known as mass precession. The time-scale to
precess by $\pi$ is given approximately by (e.g., \citealt[S4.4.1]{2013degn.book.....M})
\begin{align}
  \label{eq:tmp}
  t_\mathrm{MP} \approx \frac{1}{2} {\left (1-e^2 \right )}^{-1/2} \frac{M_\bullet}{M_\star(a)} P.
\end{align}
Mass precession changes the (in-plane) orientation of all orbits around the
MBH, which is important for RR (see \S~\ref{sect:time_scales:RR}). In the case
of a binary around the MBH, mass precession acts on the outer orbit, and this
can potentially quench LK oscillations. However, the quadrupole-order secular
MBH+binary equations of motion do not depend on $\omega_\oout$, the outer orbit
argument of periapsis. As shown below, in situations where LK-induced mergers
can potentially be enhanced due to VRR, the quadrupole-order terms dominate
(i.e., the octupole parameter $\epsilon_\mathrm{oct}$ is small, see also
\S~\ref{sect:time_scales:LK}). Therefore, mass precession is not an important
quenching source for LK oscillations in our case.

\subsection{Relativistic precession}
The lowest-order relativistic effect (assuming non-spinning MBHs) is precession
of the argument of periapsis of both inner and outer orbits, with time-scales
to precess by $\pi$ given by \citep{1972gcpa.book.....W}
\begin{subequations}
\begin{align}
  \label{eq:PN_in} t_\mathrm{1PN,\,in} &= \frac{1}{6} \left(1-e_\oin^2 \right ) \frac{a_\oin c^2}{G (M_1+M_2)} P_\oin; \\
  \label{eq:PN_out} t_\mathrm{1PN,\,out} &= \frac{1}{6} \left(1-e_\oout^2 \right ) \frac{a_\oout c^2}{G M_\bullet } P_\oout.
\end{align}
\end{subequations}
In equation~\eqref{eq:PN_out}, we assumed that $M_\bullet \gg M_1,M_2$. Also,
we neglect any PN coupling between the inner and outer orbits
\citep{2013ApJ...773..187N}. Relativistic precession in the inner orbit can be
an important source of quenching of LK cycles, and this needs to be taken into
account. Relativistic precession in the outer orbit is typically unimportant
for our purposes, for the same reason discussed in
\S\,\ref{sect:time_scales:MP}.

\subsection{Resonant Relaxation}
\label{sect:time_scales:RR}
As mentioned in the Introduction, RR arises from the torques
generated on a test orbit by correlated encounters with other stars close to
the MBH\@. One can distinguish between two regimes: the coherent regime, in which
the stellar background potential is roughly fixed (in an orbit-averaged sense),
and the incoherent regime, in which both in-plane and out-of-plane precession
cause the background potential to evolve. In the coherent regime, the
(approximately constant) net torque on a test orbit is
$\sim \sqrt{N_\star} G m_\star/a$, which leads to a change of the orbital
angular momentum
\begin{align}
\label{eq:Delta_L_coh}
  {\left ( \frac{\Delta J}{J_\mathrm{c}} \right)}_\mathrm{coh} \approx 
  \beta \frac{m_\star}{M_\bullet} \sqrt{N_\star} \frac{\Delta t}{P},
\end{align}
where $\beta$ is a dimensionless number, which can be computed as a function of
 eccentricity and density slope $\gamma$ \citep{2015MNRAS.448.3265K}. Here, for 
 simplicity, we set $\beta$ to $\beta_\mathrm{v}=2$ for VRR. One can define the VRR (or `2d RR', see also~\citealt[Section
5.6.1.3]{2013degn.book.....M}) time-scale as the duration for which
$\Delta J_\mathrm{coh} = J_\mathrm{c}$, i.e.,
\begin{align}
t_\mathrm{VRR} \equiv \frac{P}{\beta_\mathrm{v}} \frac{M_\bullet}{m_\star} \frac{1}{\sqrt{N_\star}}.
\end{align}
On time-scales longer than the coherence time-scale $t_\mathrm{coh}$, the
angular momentum of the test orbit evolves approximately as a random walk with
a step size given by the angular momentum-change accumulated during the
coherent regime, i.e.,
\begin{align}
\label{eq:Delta_L_incoh}
  {\left ( \frac{ \Delta J}{J_\mathrm{c}} \right )}_\mathrm{incoh} \approx {\left ( \frac{\Delta t}{ t_\mathrm{coh} } \right )}^{1/2}  {\left ( \frac{ \Delta J}{J_\mathrm{c}} \right )}_{\mathrm{coh,\,step}} \equiv {\left ( \frac{\Delta t}{t_\mathrm{SRR}} \right )}^{1/2},
\end{align}
where ${(\Delta J/J_\mathrm{c})}_\mathrm{coh, \,step}$ is
equation~(\ref{eq:Delta_L_coh}) evaluated at $\Delta t = t_\mathrm{coh}$, and
where after the second equality we defined the SRR (or `3d RR', see
also~\citealt[Section 5.6.1.3]{2013degn.book.....M}) time-scale. From
equations~\eqref{eq:Delta_L_coh} and~\eqref{eq:Delta_L_incoh}, it is
straightforward to show that
\begin{align}
  t_\mathrm{SRR} \equiv \frac{1}{\beta_\mathrm{s}^2} {\left ( \frac{M_\bullet}{m_\star} \right )}^2 \frac{1}{N_\star} \frac{P^2}{t_\mathrm{coh}},
\end{align}
where we set\footnote{Strictly speaking, $\beta_\mathrm{v}$ and $\beta_\mathrm{s}$ 
are not the same. However, they differ little, e.g., \citet[S5.6.1.3]{2013degn.book.....M}, and for simplicity we set
$\beta_\mathrm{s}=\beta_\mathrm{v}=2$ for the purposes of the time-scales shown in \F\,\ref{fig:timescales}.} $\beta_\mathrm{s}=2$. If the reorientation of the (orbit-averaged)
stellar background potential arises from mass precession, then
$t_\mathrm{coh} = t_\mathrm{MP}$, and (averaging $t_\mathrm{MP}$ over a thermal
distribution, \citealt{1919MNRAS..79..408J})
\begin{align}
  t_\mathrm{SRR} = \frac{1}{\beta_\mathrm{s}^2} \frac{M_\bullet}{m_\star} P,
\end{align}
such that $t_\mathrm{VRR} = (\beta_\mathrm{s}^2/\beta_\mathrm{v}) \, t_\mathrm{SRR} \, \sqrt{N_\star}$. This shows
that $t_\mathrm{VRR} < t_\mathrm{SRR}$, unless $N_\star$ is small (i.e., very
close to the MBH).

Coherence of the stellar background is generally broken by relativistic
precession or mass precession, whichever process dominates. We write the
effective coherence time-scale as
\begin{align}
  \label{eq:t_coh}
  t_\mathrm{coh}^{-1} = \langle \langle t_\mathrm{1PN,\, out} \rangle \rangle^{-1} + \langle \langle t_\mathrm{MP} \rangle \rangle^{-1},
\end{align}
where the double brackets denote averages over $e_\oout$ assuming a thermal
distribution.

\subsection{Nonresonant relaxation}
\label{sect:time_scales:NRR}
At larger distances from the MBH, encounters become uncorrelated, and RR
becomes ineffective. Uncorrelated encounters give rise to nonresonant
relaxation (NRR), which changes all orbital elements of a test orbit on a
time-scale given by
\begin{align}
  \label{eq:NRR}
  t_\mathrm{NRR} \approx C_\mathrm{NRR} (\gamma) {\left ( \frac{M_\bullet}{m_\star} \right )}^2 \frac{P}{N_\star \ln \Lambda}.
\end{align}
Here, the NRR time-scale is defined in terms of the angular-momentum diffusion
coefficient in the limit of zero angular-momentum; the dimensionless quantity
$C_\mathrm{NRR}(\gamma)$ is a function of the density slope $\gamma$ (see
Appendix B of \citealt{2014MNRAS.443..355H}). For the Coulomb logarithm, we
assume $\Lambda = M_\bullet/(2m_\star)$. From equation~(\ref{eq:NRR}), it is
clear that $t_\mathrm{NRR} \gg t_\mathrm{VRR}$.

\subsection{Binary evaporation}
\label{sect:time_scales:EV}
Uncorrelated encounters with other stars not only affect the orbital elements
of the outer orbit, but also those of the inner orbit \citep{2008gady.book.....B,2009ApJ...690..795P}.
The nature of the effect
on the inner orbit semimajor axis depends on the ratio of the (absolute value
of the) binding energy, $\Ec_\oin=G(M_1+M_2)/(2a_\oin)$, to the squared stellar
velocity dispersion, $\sigma_\star^2(r)$ \citep{1975MNRAS.173..729H}. If the
inner binary is soft, $G(M_1+M_2)/(2a_\oin) \ll \sigma_\star^2$, then
encounters tend to make it even softer ($a_\oin$ increases), whereas hard
binaries, for which $G(M_1+M_2)/(2a_\oin) \gg \sigma_\star^2$, tend to become
harder ($a_\oin$ decreases; e.g.,
\citealt{1993ApJ...403..256H,1996ApJ...467..359H}).

For soft binaries, the time-scale for the inner orbit binding energy to change
by order itself is given by
\citep{2014ApJ...780..148A}
\begin{align}
  \label{eq:t_EV}
  t_\mathrm{EV} = \frac{1}{8} \sqrt{ \frac{1+q_\sigma}{2\pi q_\sigma}} \frac{M_1+M_2}{m_\star} \frac{\sigma_\star(r)}{G n_\star(r) m_\star a_\oin \ln \Lambda_\oin},
\end{align}
where $q_\sigma \equiv (M_1+M_2)/m_\star$, and
$\Lambda_\oin = 3(1+1/q_\sigma)/(1+2/q_\sigma) [ \sigma_\star^2(r)/v_\oin^2]$,
with $v_\oin^2 = G(M_1+M_2)/a_\oin$ \citep{2014ApJ...780..148A}.

\subsection{LK cycles}
\label{sect:time_scales:LK}
The quadrupole-order time-scale of LK oscillations can be estimated as (e.g.,
\citealt{1997AJ....113.1915I,1999CeMDA..75..125K,2015MNRAS.452.3610A})
\begin{align}
  \label{eq:t_LK}
  t_\mathrm{LK} = \frac{8}{15\pi} \left ( \frac{P_\oout^2}{P_\oin} \right) \frac{M_1+M_2+M_\bullet}{M_\bullet} {\left (1-e_\oout^2 \right )}^{3/2}.
\end{align}
We shall refer to the ratio of the LK and VRR time-scales as the `adiabatic
parameter' $\adpar$,
\begin{align}
\label{eq:adpar}
  \adpar & 
           \equiv
           \frac{t_\mathrm{LK}}{t_\mathrm{VRR}} = \frac{8}{15\pi}
           \frac{P_\oout}{P_\oin}  \frac{M_1+M_2+M_\bullet}{M_\bullet}
           \frac{m_\star}{M_\bullet} \beta_\mathrm{v} \sqrt{N_\star} {\left (1-e_\oout^2
           \right )}^{3/2} 
           \nonumber \\
         & 
           \simeq
           \frac{8}{15 \pi} {\left ( \frac{a_\oout}{a_\oin} \right )}^{3/2} {\left ( \frac{M_1+M_2}{M_\bullet} \right )}^{1/2} \frac{m_\star}{M_\bullet} \beta_\mathrm{v} \sqrt{N_\star} {\left (1-e_\oout^2 \right )}^{3/2},
\end{align}
where in the last line we assumed $M_\bullet \gg M_1,M_2$. One can consider
$\adpar \ll 1$ to be the `adiabatic' regime in which VRR slowly changes the
orientation of the outer orbit, and on time-scales $\ll t_\mathrm{VRR}$, the
system is well described as an isolated three-body system. Non-adiabatic
regimes ($\adpar \sim 1$ or $\adpar \gg 1$) are associated with different
behavior, and we consider these regimes in more detail in \S\,\ref{sect:dyn}.

The importance of the octupole-order terms can be estimated by the octupole
parameter $\epsilon_\mathrm{oct}$ (which is essentially the ratio of the
coefficient of the octupole-order term to the coefficient of the
quadrupole-order term),
\begin{align}
  \label{eq:eps_oct}
  \epsilon_\mathrm{oct} \equiv \frac{|M_1-M_2|}{M_1+M_2} \frac{a_\oin}{a_\oout} \frac{e_\oout}{1-e_\oout^2}.
\end{align}
Orbital flips, and associated high eccentricities, can occur if
$\epsilon_\mathrm{oct} \gtrsim 10^{-3}$
\citep{2011ApJ...742...94L,2011PhRvL.107r1101K,2013ApJ...779..166T,2014ApJ...791...86L}. We
evaluate typical values of $\epsilon_\mathrm{oct}$ for our systems of interest
below.

\subsection{Comparison of times-scales for different parameters}
\label{sect:time_scales:comp}
We compute the various time-scales discussed above for clusters with different
$M_\bullet$ and $\gamma$, and for different inner binary semimajor axes
$a_\oin$. We set the inner binary masses to $M_1 = 30\,\msun$ and
$M_2 = 20\, \msun$, representing some of the recent LIGO systems, and the
outer orbit eccentricity is set to $e_\oout=2/3$ (the mean value for a thermal
distribution, \citealt{1919MNRAS..79..408J}). The inner orbit eccentricity,
used to calculate $t_\mathrm{1PN,\,\oin}$, is set to $e_\oin=0$ (an optimistic
assumption in terms of quenching of LK oscillations). The average background
stellar mass is set to $m_\star = 1\,\msun$.

To set the overall normalization of the stellar distribution, we adopt the
$M_\bullet-\sigma_\mathrm{bulge}$-relation between the MBH mass and bulge
velocity dispersion $\sigma_\mathrm{bulge}$ \citep{2001ApJ...547..140M},
\begin{align}
\log_{10} \left ( \frac{M_\bullet}{\msun} \right ) = 4.8 \log_{10} \left ( \frac{\sigma_\mathrm{bulge}}{\mathrm{km\,s^{-1}}} \right ) - 2.9,
\end{align}
and set $\sigma_\mathrm{bulge}$ to $\sigma_\star(r_\mathrm{h})$, where
$r_\mathrm{h}$ is the radius at which the enclosed distributed stellar mass is
$2M_\bullet$ (i.e., combining two definitions for the radius of influence).

Specifically, assuming a spherically symmetric stellar number density
distribution $n_\star(r) = n_\mathrm{h} {(r/r_\mathrm{h})}^{-\gamma}$ and using
the isotropic Jeans equation, one can show that $r_\mathrm{h}$ and
$\sigma_\star$, where $\sigma_\star$ is the velocity dispersion at $r=r_\mathrm{h}$,
 are related via
\begin{align}
\label{eq:r_h_from_sigma}
  r_\mathrm{h} = \frac{GM_\bullet}{\sigma^2_\star} \frac{1}{1+\gamma} \left (1 + \frac{1+\gamma}{\gamma-1} \right ).
\end{align}
For a given $M_\bullet$ and $\gamma$, we compute $\sigma_\mathrm{bulge}$ from
the $M_\bullet-\sigma_\mathrm{bulge}$-relation \citep{2001ApJ...547..140M}, set
$\sigma_\star = \sigma_\mathrm{bulge}$, and calculate
$r_\mathrm{h}$ from equation~\eqref{eq:r_h_from_sigma}. The normalization
$n_\mathrm{h}$ is then given by
$n_\mathrm{h} = [(3-\gamma)/(4\pi)] (2M_\bullet/m_\star)
r_\mathrm{h}^{-3}$. Related quantities such as $N_\star$ and $M_\star$ are
found by straightforward integration over volume of $n_\star(r)$.
 
\begin{figure*}
  \center
\includegraphics[scale = 0.45, trim = 5mm 0mm 0mm 20mm]{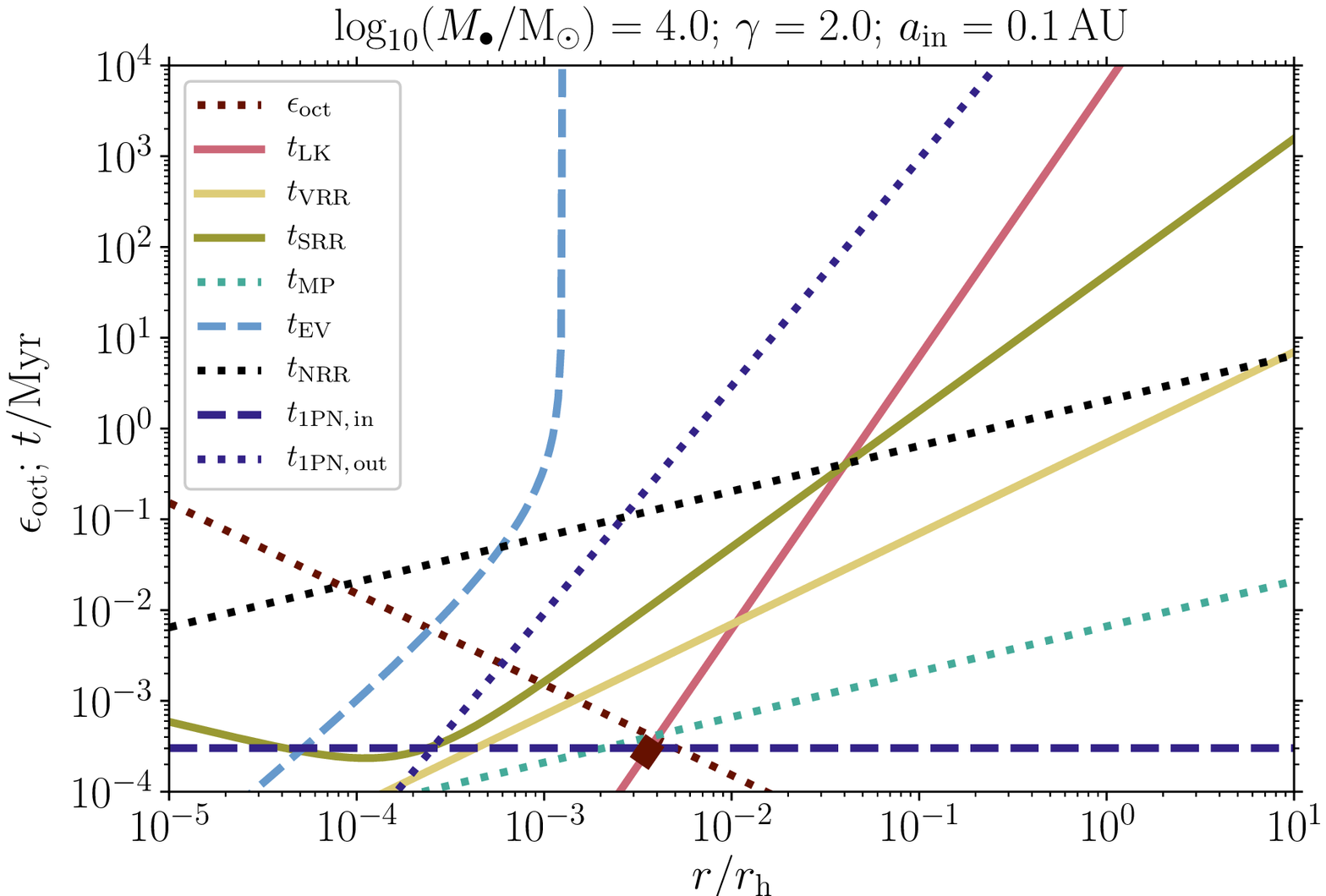}
\includegraphics[scale = 0.45, trim = 5mm 0mm 0mm 20mm]{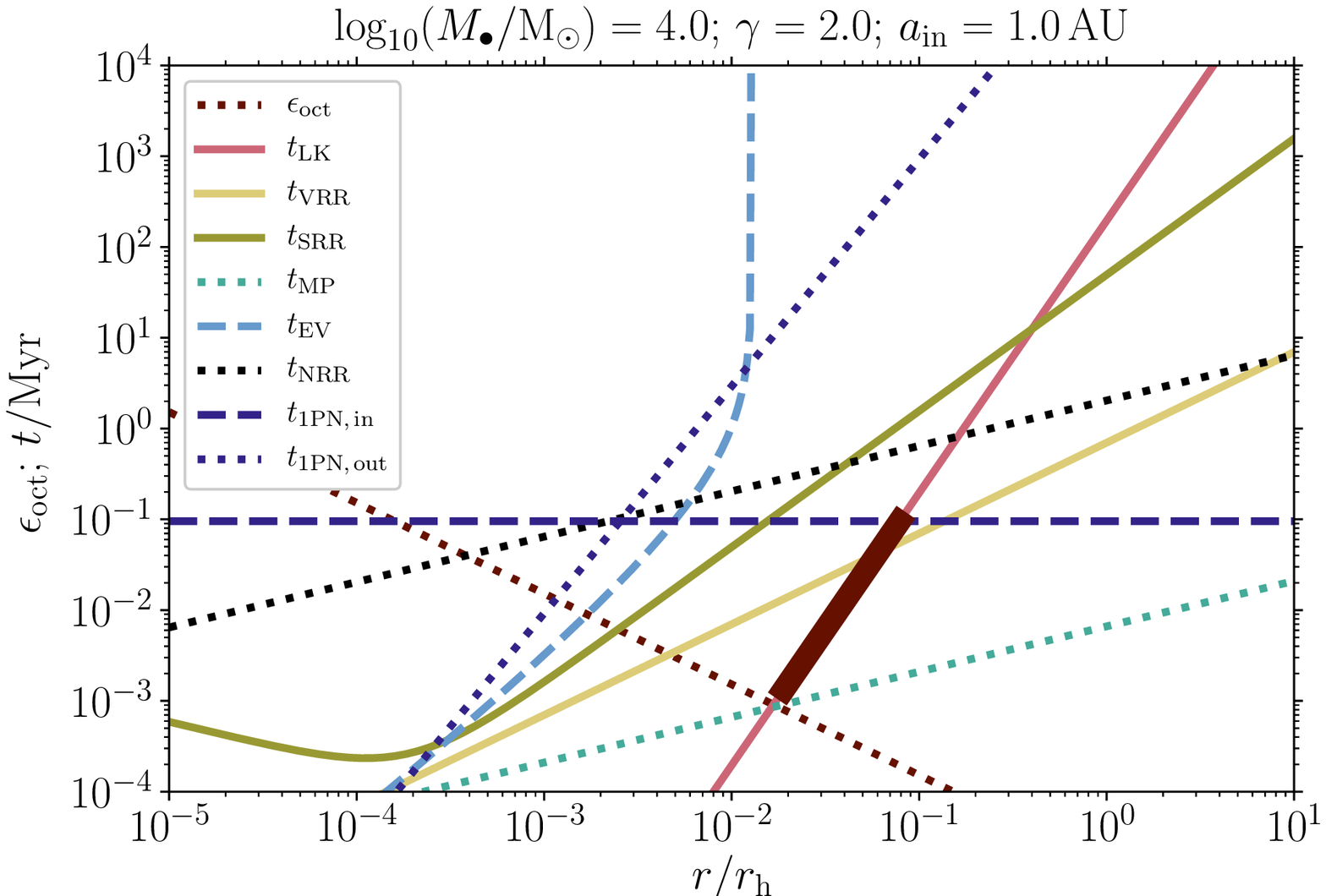}
\includegraphics[scale = 0.45, trim = 5mm 0mm 0mm 0mm]{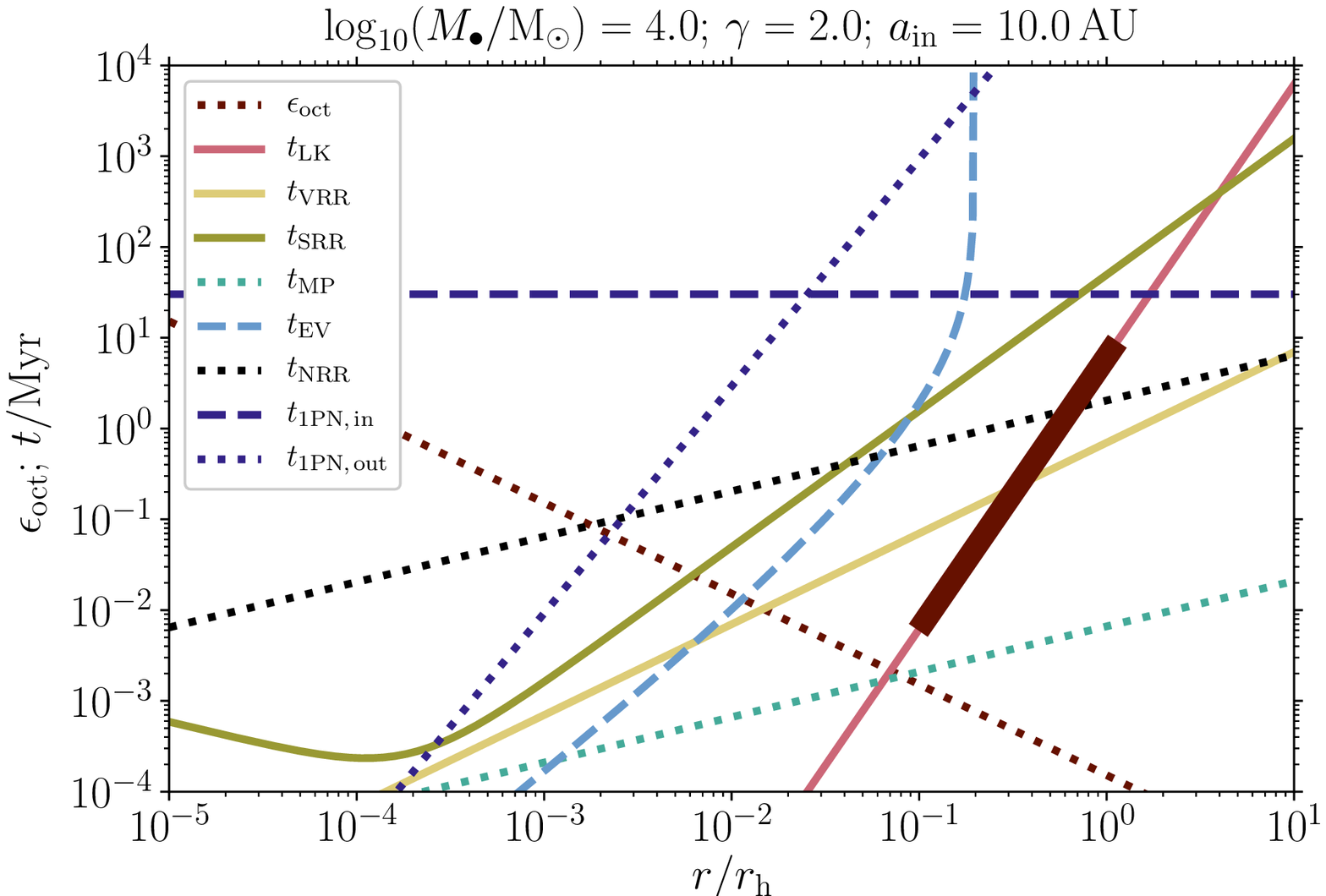}
\includegraphics[scale = 0.45, trim = 5mm 0mm 0mm 0mm]{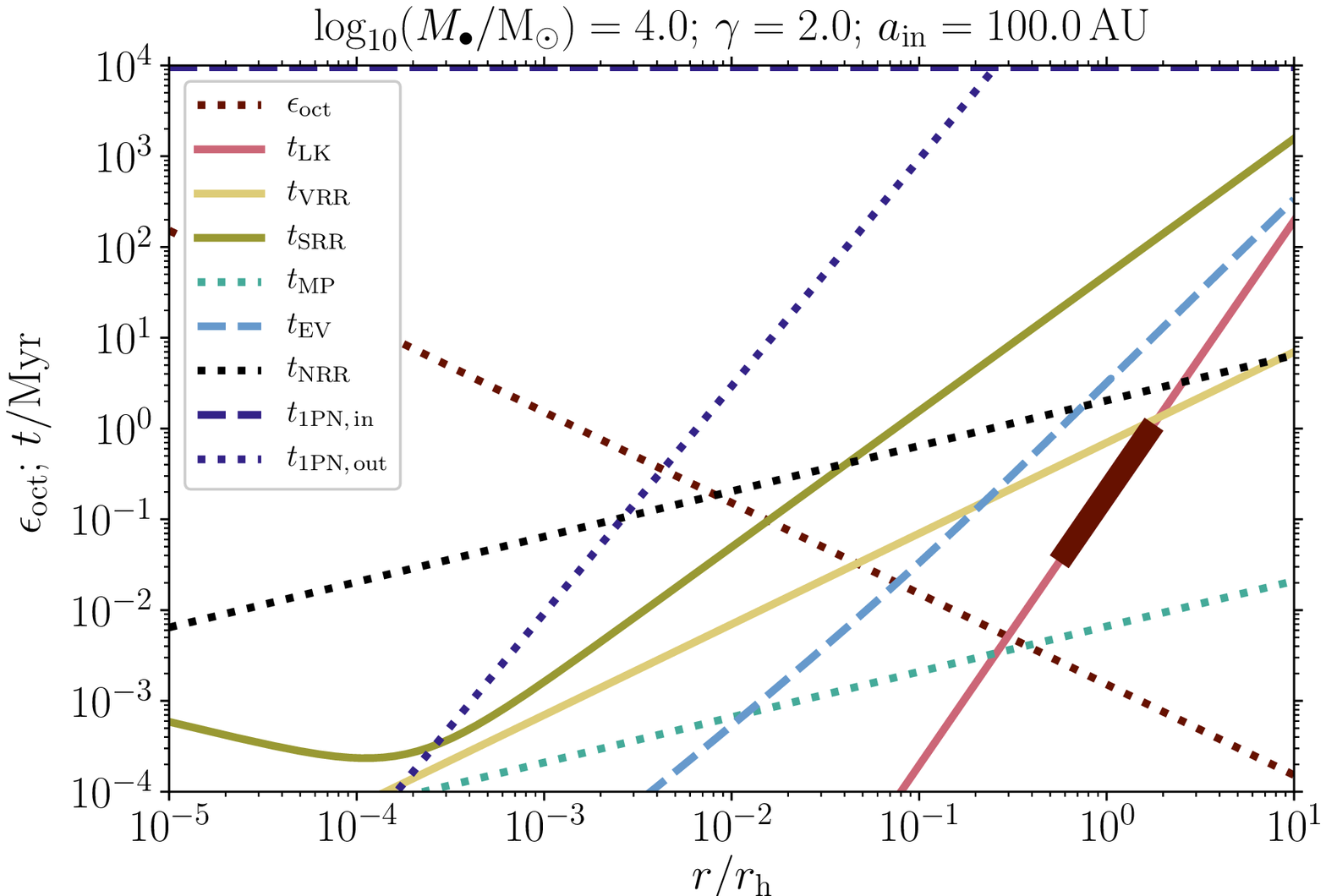}
\includegraphics[scale = 0.45, trim = 5mm 0mm 0mm 0mm]{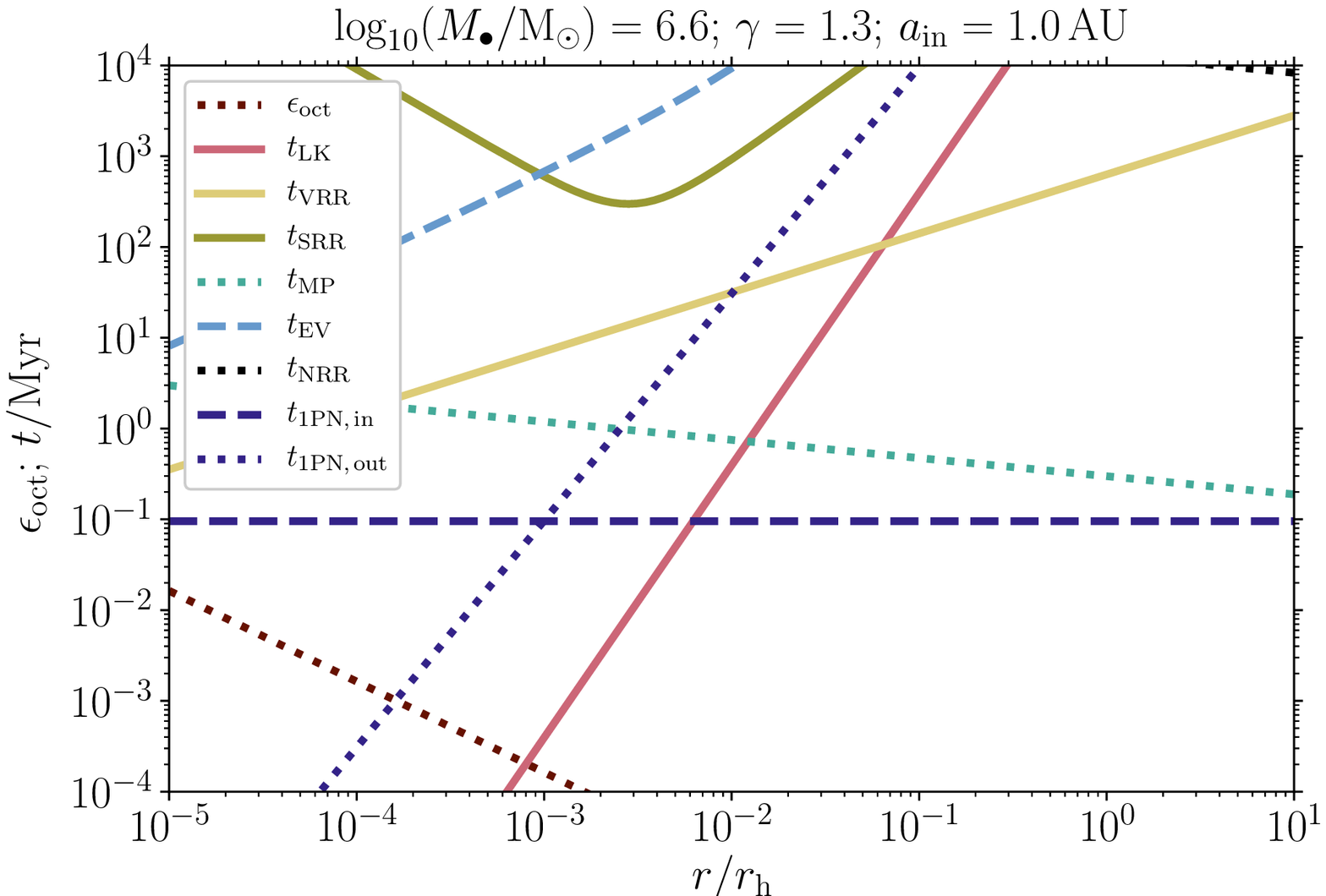}
\includegraphics[scale = 0.45, trim = 5mm 0mm 0mm 0mm]{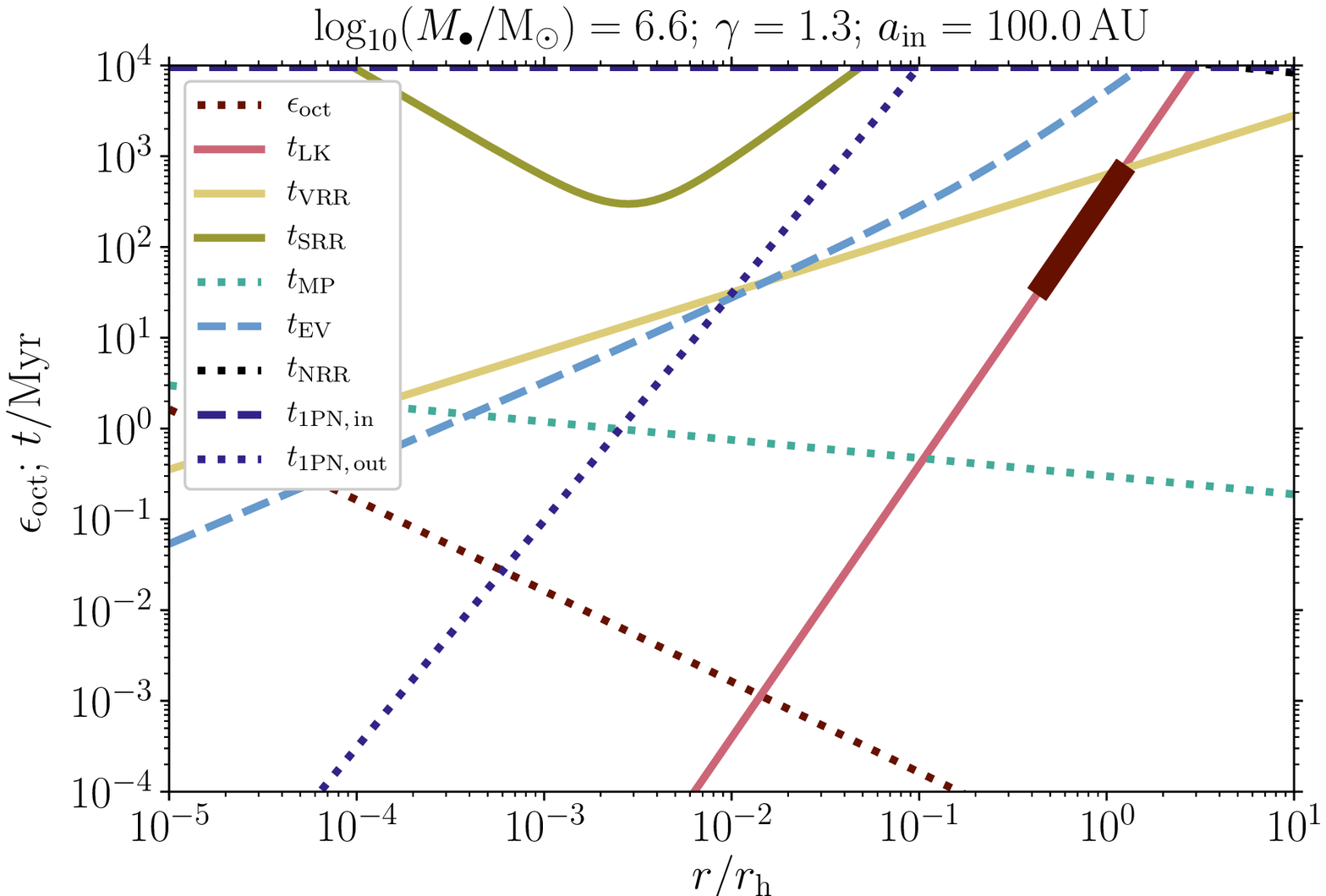}
\includegraphics[scale = 0.45, trim = 5mm 0mm 0mm 0mm]{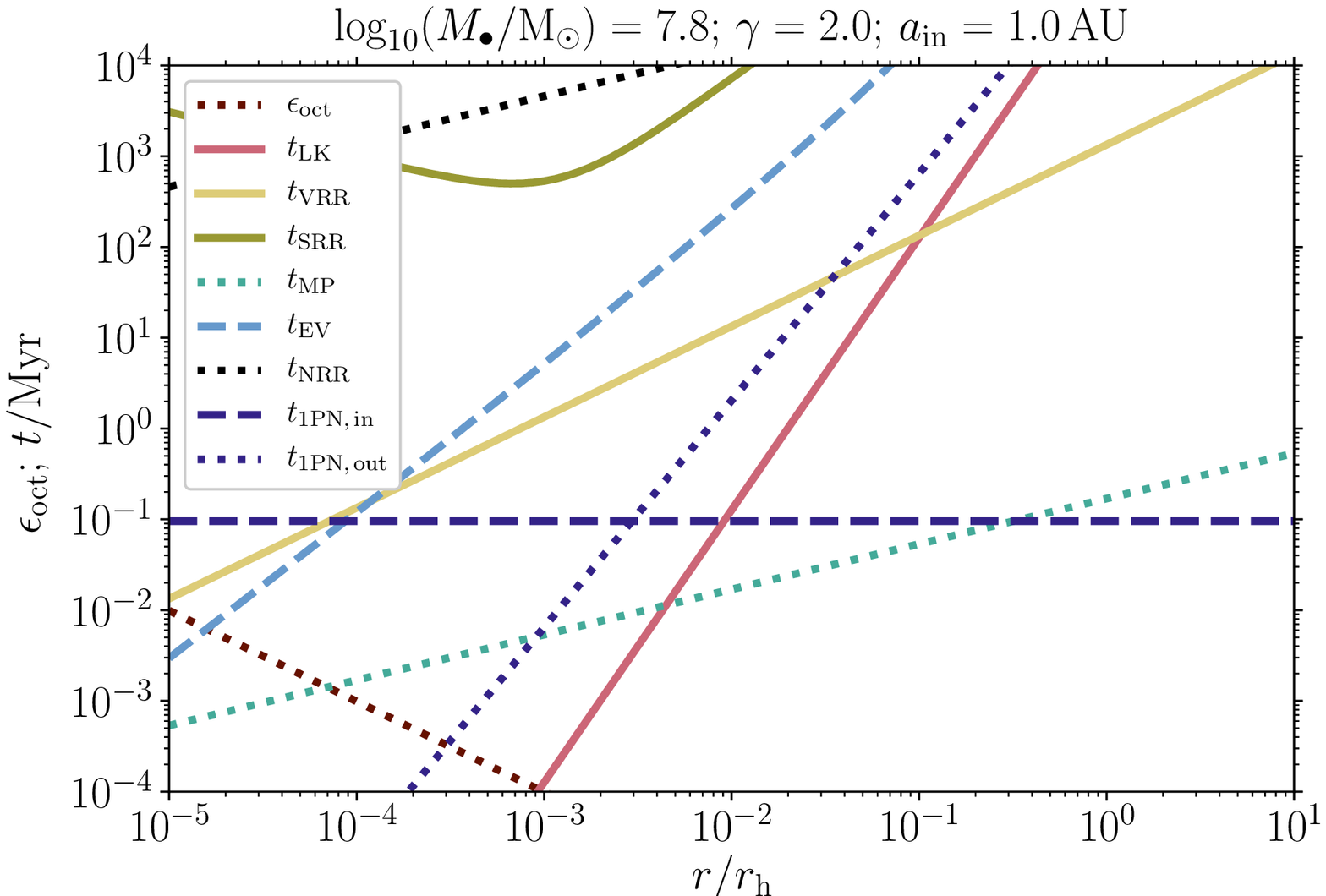}
\includegraphics[scale = 0.45, trim = 5mm 0mm 0mm 0mm]{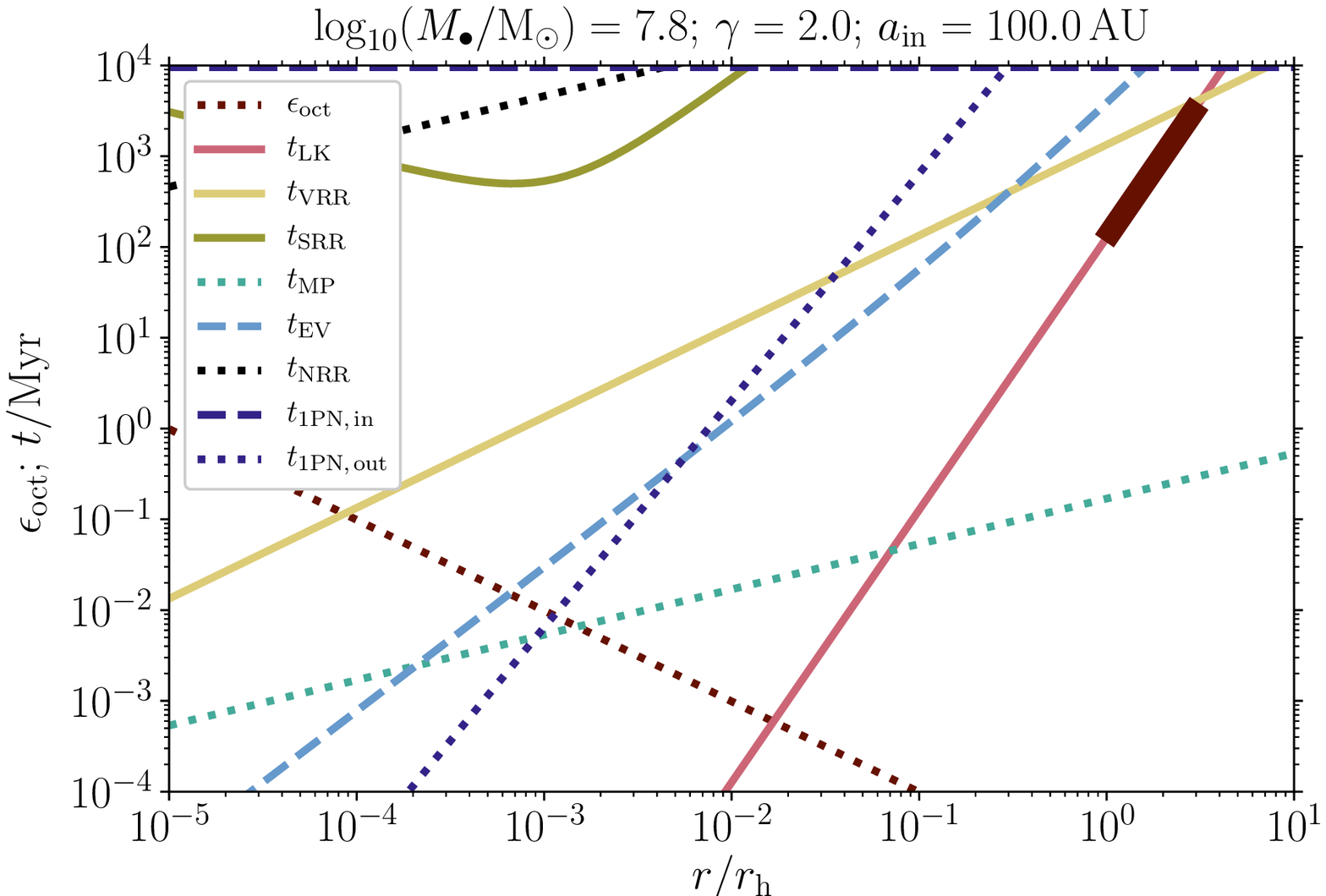}
\caption{\small Various time-scales and octupole parameter as a function of
  distance to the MBH, $r$, normalized to $r_\mathrm{h}$. The octupole
  parameter (equation~\ref{eq:eps_oct}) is shown with the dark red dotted
  line. Other lines show various time-scales: $t_\mathrm{LK}$: red solid;
  $t_\mathrm{VRR}$: light yellow solid; $t_\mathrm{SRR}$: dark yellow solid;
  $t_\mathrm{MP}$: light green dotted; $t_\mathrm{EV}$: light blue dashed;
  $t_\mathrm{NRR}$: black dotted; $t_\mathrm{1PN,\,in}$: dark blue dashed (setting $e_\oin=0$);
  $t_\mathrm{1PN,\,out}$: dark blue dashed. Each panel corresponds to certain
  values of $M_\bullet$, $\gamma$, and $a_\oin$, indicated in the top. The
  thick solid dark red lines indicate the radii for which we expect VRR to be
  important (not applicable in all panels). }
\label{fig:timescales}
\end{figure*}

We show the resulting time-scales and $\epsilon_\mathrm{oct}$ as a function of
$r$, normalized by $r_\mathrm{h}$, in \F\,\ref{fig:timescales}. Each panel
corresponds to a different combination of $M_\bullet$, $\gamma$ and
$a_\oin$. Refer to the legends and/or the figure caption for the meaning of the
various colors and line styles.

The regions of interest are where the LK and VRR time-scales are similar, and a
number of other conditions are met. In \F\,\ref{fig:timescales}, we highlight
with thick solid dark red lines the radii for which
\begin{enumerate}[(i)]
\item $0.1 < \adpar< 10$;
\item $t_\mathrm{LK} < t_\mathrm{EV}/10$;
\item $t_\mathrm{LK} < 10 \, \mathrm{Gyr}$;
\item $t_\mathrm{LK} < t_\mathrm{1PN,\,in}$.
\end{enumerate}
The chosen range for $\adpar$ in condition (i) is somewhat arbitrary. We
investigate the dynamics as a function of $\adpar$ in more detail in
\S\,\ref{sect:dyn}. Here, the main aim is to broadly explore the parameter
space and pinpoint regions of interest, rather than to accurately capture the
dynamics. Condition (ii) ensures that at least $\mathcal{O}(10)$ LK cycles
occur before the binary evaporates. Condition (iii) excludes LK time-scales
that are too long compared to the age of the Universe, and condition (iv)
excludes systems in which LK cycles are suppressed by relativistic precession
in the inner orbit.

We note the following:
\begin{itemize}
\item There are no regions of interest for tight binaries
  ($a_\oin=0.1\,\au$). This can be ascribed to the short relativistic
  precession time-scale.
\item The kink as a function of $r$ in the SRR time-scale is due to a change of
  1PN precession being the dominant reshuffling process close to the MBH, to
  mass precession further away (see equation~\ref{eq:t_coh}).
\item SRR can be as important as VRR close to the MBH, but the regions of
  interest are located further from the MBH, where
  $t_\mathrm{SRR} \gg t_\mathrm{VRR}$.
\item The effect of binary evaporation is weak for the regions of interest.
\item The NRR time-scale is typically longer than the LK and VRR
  time-scales in the regions of interest, although there are some cases in
  which they are comparable (in particular, for low MBH masses and large
  $a_\oin$).
\item In the regions of interest, the octupole parameter is typically small
  ($\epsilon_\mathrm{oct} \lesssim 10^{-4}$), indicating that the eccentric LK
  mechanism is expected to be unimportant. Nevertheless, we include the
  octupole (and higher-order) terms in the numerical integrations in
  Sections~\ref{sect:dyn} and~\ref{sect:pop_syn} (see also \S~\ref{sect:meth:sec}).
\end{itemize}

\subsection{Potentially interesting regimes}
\label{sect:time_scales:int}
Although \F~\ref{fig:timescales} contains detailed information on the
time-scales, it is difficult to obtain insight into the dependence of the size
of the parameter space of interest as a function of $M_\bullet$, $\gamma$ and
$a_\oin$. For that purpose, we here determine the number of stars within the
radial extent of interest, normalized to the total number of stars. We consider
ranges in $r$ between $10^{-5}$ and $10\,r_\mathrm{h}$.

Specifically, let $r_\mathrm{in}$ and $r_\mathrm{out}$ denote the inner and
outer edges of the region for which the criteria discussed in
\S~\ref{sect:time_scales:comp} are satisfied (i.e., the radial region indicated
with the thick solid dark red lines in \F~\ref{fig:timescales}). The number of
stars within the region of interest is then
$N_\mathrm{interest} = N_\star(r_\mathrm{out}) - N_\star(r_\mathrm{in})$; the
total number of stars is $N_\mathrm{tot} = N_\star(10 \,
r_\mathrm{h})$. Evidently, $N_\mathrm{interest}$ does not give the actual
number of {\it binaries} of interest; to get the latter number, one should
multiply by the binary fraction. We emphasize that
$N_\mathrm{interest}/N_\mathrm{tot}$ does not contain direct information on the
enhancement of merger rates due to VRR\@; it measures the importance of the
parameter space for which LK-VRR coupling is potentially important.

\begin{figure}
\center
\includegraphics[width=0.48\textwidth]{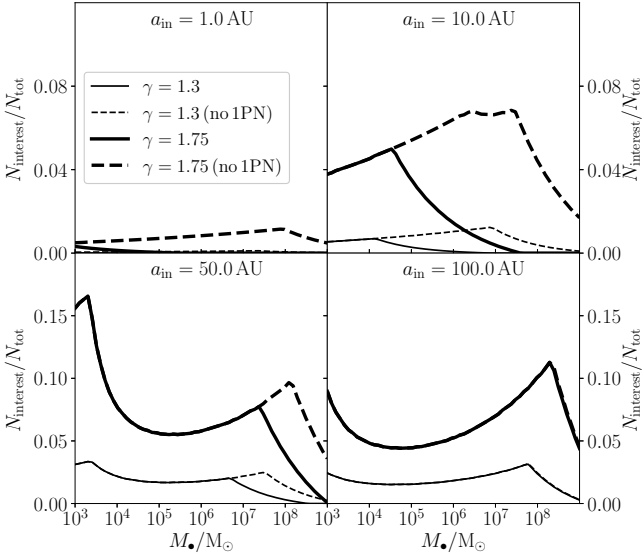}
\includegraphics[width=0.48\textwidth]{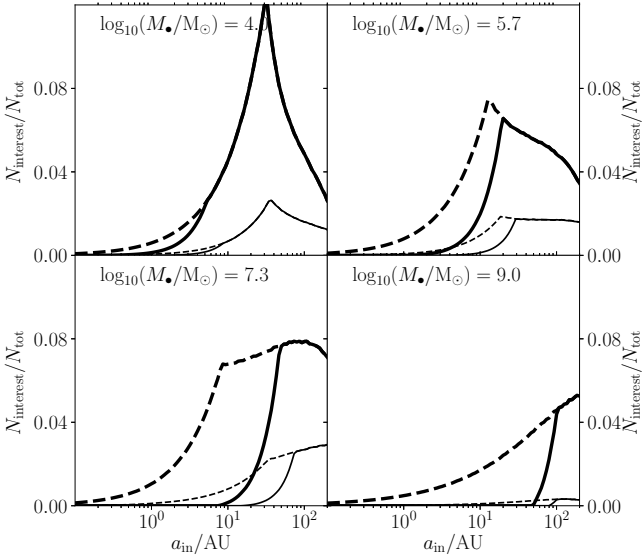}
\caption{\small Number of stars within the region of interest,
  $N_\mathrm{interest}$, normalized to the total number of stars,
  $N_\mathrm{tot}$ (see text in \S~\ref{sect:time_scales:int}). The region of
  interest is defined with the criteria in \S~\ref{sect:time_scales:comp}. The
  ratios $N_\mathrm{interest}/N_\mathrm{tot}$ are shown as a function of
  $M_\bullet$ for several values of $a_\oin$ and $\gamma$ (top four panels),
  and as a function of $a_\oin$ for several values of $M_\bullet$ and $\gamma$
  (bottom four panels). Solid (dashed) lines show
  $N_\mathrm{interest}/N_\mathrm{tot}$ with (without) taking into relativistic
  precession in the inner orbit. }
\label{fig:regime_fractions}
\end{figure}

In \F~\ref{fig:regime_fractions}, we show $N_\mathrm{interest}/N_\mathrm{tot}$
as a function of $M_\bullet$ for several values of $a_\oin$ and $\gamma$ (top
four panels), and as a function of $a_\oin$ for several values of $M_\bullet$
and $\gamma$ (bottom four panels). To illustrate the importance of relativistic
precession in the inner binary, we show results with (solid lines) and without
(dashed lines) the inclusion of criterion (iv) described in
\S~\ref{sect:time_scales:comp}.

From \F~\ref{fig:regime_fractions}, it is clear that there is a strong
dependence of $N_\mathrm{interest}/N_\mathrm{tot}$ on all three parameters
$M_\bullet$, $\gamma$ and $a_\oin$. Generally, the number fraction is
significant (i.e., $\gtrsim 0.04$) for a relatively narrow range of $a_\oin$,
which can be understood by noting that relativistic precession dominates in
tight binaries, whereas evaporation and the age of the Universe restrict the
region of interest for wide binaries. More massive MBHs imply longer VRR
time-scales whereas the LK time-scale decreases. Therefore,
$N_\mathrm{interest}/N_\mathrm{tot}$ decreases for large values of $M_\bullet$.

We investigate the dependence of the importance of LK-VRR coupling and merger
rates on the parameters in more detail in \S~\ref{sect:pop_syn}.

\section{Methodology}
\label{sect:meth}

\subsection{Model for Vector Resonant Relaxation}
\label{sect:meth:VRR}

In order to mimic the stochastic motion of the outer orbit by VRR, we construct
a toy model which, statistically, reproduces the orbital evolution of a test
star due to the stochastic torques of the background. This toy model is based
on the assumption that the two important properties of the stochastic torques
are the rms of the torque
${\tau_\rv = J\sqrt{\langle\dot{\hat{\bJ}}\cdot\dot{\hat{\bJ}}\rangle}} =
\beta_\rv\!\sqrt{N_\star}m_\star J_\rc^2/(M_\bullet P)$ and the time scale on
which this torque remains coherent $\Tcvrr$, which is set by the VRR timescale,
$\Tcvrr = \epsilon_\rc t_\mathrm{VRR}$. The free parameter $\beta_\rv$, which
determines the magnitude of the torque, can be calculated directly
\citep[e.g.,][]{2015MNRAS.448.3265K}. To determine $\epsilon_\rc$, which sets
the coherence timescale, we fit the orbital evolution in our toy model to
numerical simulations.

We assume that the motion of a test orbit is governed by the quadruple
moment of the background stars.
The evolution of the angular-momentum vector is given by the equation of motion
\begin{equation}
  \label{eq:Jdot}
  \dot{\hat{\bJ}} = J^{-1}\tau_\rv (Q(t) \hat{\bJ}) \times \hat{\bJ},
\end{equation}
where $Q(t)$ is a time dependent matrix
\begin{equation}
  \label{eq:Q}
Q=\sqrt{\frac{3}{4}}\left(\begin{array}{ccc}
\eta_{2\rr}(t) & \eta_{2\ri}(t) & -\eta_{1\rr}(t)\\
\eta_{2\ri}(t) & -\eta_{2\rr}(t) & -\eta_{1\ri}(t)\\
-\eta_{1\rr}(t) & -\eta_{1\ri}(t) & \sqrt{3}\eta_{0}(t)
\end{array}\right),
\end{equation}
which depends on the intricate motion of the background stars through its
decomposition into spherical harmonics
${\eta_{i\rr}(t) \propto \sum_{n}\Re\big[Y_{2i}(\hat{\bJ}_n(t))\big]}$,
${\eta_{i\ri}(t) \propto \sum_{n}\Im\big[Y_{2i}(\hat{\bJ}_n(t))\big]}$, where
$\{\hat{\bJ}_n(t)\}$ are angular momentum vectors of background stars.  This
model has two free parameters, $\beta_\rv$ and $\epsilon_\rc$, which determine
the magnitude of the torque and the coherence time, respectively.

Instead of following the complex motion of the background, we replace the
$\eta_i(t) \in \{\eta_0, \eta_{1\rr}(t), \eta_{1\ri}(t), \eta_{2\rr}(t),
\eta_{2\ri}(t)\}$ functions by Gaussian noise terms with similar statistical
properties as the exact ones.

These noise terms are of zero mean ${\langle\eta_i(t) \rangle = 0}$, and are fully
described by their correlation function
${\langle\eta_i(t)\eta_j(t') \rangle = \delta_{ij}C(t-t')}$, which is
normalized such that ${ \langle\dot{\bJ}\cdot\dot{\bJ}\rangle =
  \tau_\rv^2}$. The coherence time, which is the timescale on
which $C(t-t')$ decays, is set by fitting the correlation in $\hat{\bJ}$ to
effective $N$-body simulations.

Here, we implement correlated noise with a correlation timescale
$\Tcvrr$, by drawing ${\eta_j^i}$ from a Gaussian distribution, with zero
mean and standard deviation of $1$, at times ${t_i}$ each coherence time
$t_{i+1}-t_{i} = \Tcvrr$, and linearly interpolating between
$\eta_j^i$ and $\eta_j^{i+1}$. Thus, on
the interval $t_i \le t  < t_{i+1}$, the noise terms are given by
\begin{equation}
  \label{eq:eta_t}
  {\eta_j(t) = \eta_j^{i} + \left(\eta_j^{i+1} - \eta_j^i \right )(t-t_i)/\Tcvrr},
\end{equation}
and the correlation function is
\begin{equation}
  \label{eq:corr}
  C(t) = \begin{cases}
    \displaystyle
    \frac{2}{3} + \frac{t-2 \Tcvrr}{2{(\Tcvrr)}^3} t^2,    &  |t| \le \Tcvrr; \\
    \displaystyle
  \frac{{(2\Tcvrr - |t|)}^3}{6 {(\Tcvrr)}^3},  &  \Tcvrr < |t| \le 2\Tcvrr; \\ 
0, & |t| > 2\Tcvrr.
\end{cases}
\end{equation}
To compare this scheme to numerical simulations, we define the mean squared
change in orientation of the angular momentum over time ${\Delta t = t' -t}$ as
\begin{equation}
  \label{eq:dJ2}
  \left \langle {(\Delta \hat{\bJ})}^2  \right \rangle
  \equiv
  \frac{1}{2} \bigg\langle 
  {\big| \hat{\bJ}(t) -  \hat{\bJ}(t') \big|}^2
  \bigg\rangle
  = 1 -  \left \langle \hat{\bJ}(t)\cdot\hat{\bJ}(t')  \right \rangle.
\end{equation}
On short timescales, ${t\ll \Tcvrr}$, the torque is nearly constant and the
evolution of $\bJ$ is ballistic
$ \left \langle {(\Delta \hat{\bJ})}^2  \right  \rangle \approx \tau_\rv
t^2/(2J)$. On long timescales, ${t\gg \Tcvrr}$, the torque is
reshuffled and $\bJ$ changes in a diffusive manner $\left \langle {(\Delta
  \hat{\bJ})}^2 \right \rangle \approx 1-\exp(-D_\mathrm{VRR}t)$, where
$D_\mathrm{VRR} = J^{-2} \tau_\rv^2 {(\Tcvrr)}^2$ is the
diffusion coefficient.

\begin{figure}
\center
\includegraphics[width=0.48\textwidth]{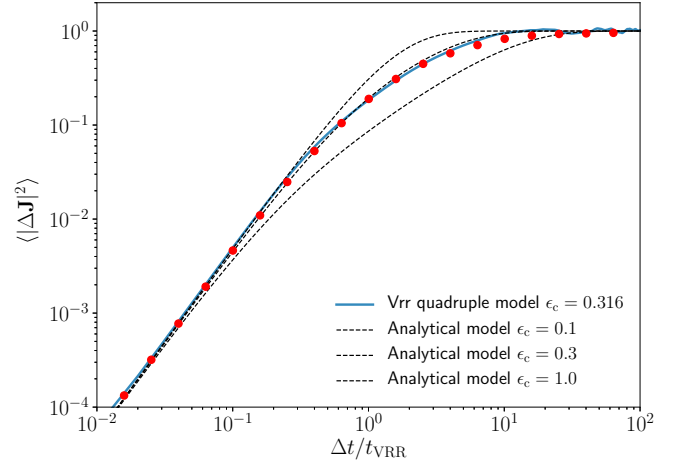}
\caption{Change in the direction of the angular-momentum of the outer orbit ${\left \langle
    {(\Delta \hat{\bJ})}^2 \right \rangle}$ as a function of time for the VRR
  model (solid blue line) fitted to the data from numerical simulations of interacting rings (red
  dots). For reference, the black dashed lines show the analytic approximation in
  equation~(\ref{eq:dJ}) with different values of $\epsilon_\rc$. 
\label{fig:autocorrelations}}
\end{figure}

We verified our model for VRR by carrying out a number of integrations of the
equations of motion (without the terms accounting for the secular binary+MBH
interaction), and by comparing the evolution of $\hat{\bJ}$ to numerical
simulations of interacting rings. In these simulations, we used a set of
gravitationally interacting circular rings and advanced the system by
calculating the mutual torques between all ring-pairs at each
time-step~\citep[see, e.g.,][]{Ulubay-Siddiki+2009}.
As shown in Fig~\ref{fig:autocorrelations},
$ \left \langle {(\Delta \hat{\bJ})}^2 \right \rangle$ can be analytically approximated by
\begin{equation}
  \label{eq:dJ}
  \left \langle \hat{\bJ}(0)\cdot \hat{\bJ}(t)
  \right \rangle
  \approx 1 - \exp \left [ -J^{-2}\tau_\rv^2 {(\Tcvrr)}^2 \chi(t/\Tcvrr) \right ],
\end{equation}
where ${\chi(x) = x - 1 + \exp(-x)}$, and the model is consistent with the ring
simulations, for ${\epsilon_\rc \approx 0.316}$.

\subsection{Secular dynamics}
\label{sect:meth:sec}
To describe the secular dynamics of the binary+MBH system, we adopt the
standard equations of motion for hierarchical triple systems based on an
expansion of the Hamiltonian in terms of the ratio of the separations of the
inner and outer orbits. The Hamiltonian is averaged over both orbits, and the
equations of motion (in vector form) are integrated numerically. By default, we
include the quadrupole, octupole, hexadecapole and dotriacontupole-order terms
as implemented within \textsc{SecularMultiple} \citep{2016MNRAS.459.2827H}, in
which we have included the VRR model described in
\S~\ref{sect:meth:VRR}. Although not studied here, the implementation also
supports the inclusion of VRR on the orbits of more complicated subsystems
orbiting the MBH, e.g., multiplanet systems within binaries.

\section{Understanding the dynamics}
\label{sect:dyn}
In this section, we address in detail the impact of VRR on the LK evolution of
the binary+MBH system. We consider the VRR and LK dynamics only, i.e., we
include the LK equations of motion and the effects of VRR on the outer
orbit. Other physical processes such as PN corrections and binary evaporation
are ignored here, but are included in the Monte Carlo integrations in
\S~\ref{sect:pop_syn}.

\subsection{Illustration of the different regimes}
\label{sect:dyn:ex}

\begin{figure}
\center
\includegraphics[width=0.52\textwidth]{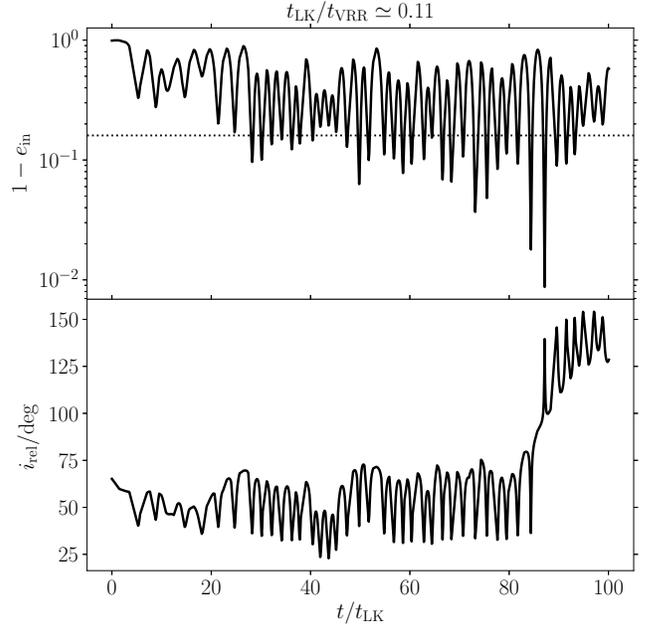}
\caption{\small Example evolution in the `mildly transadiabatic' regime,
  $\mathcal{R}\simeq 0.11$. Top panel: inner orbit eccentricity; bottom panel:
  mutual inclination between the inner and outer orbits. The initial mutual
  inclination is $i_\mathrm{rel,\,0} \simeq 65^\circ$, and in the canonical
  case without VRR, the maximum eccentricity (see equation~\ref{eq:e_max_LK})
  is $\simeq 0.84$ (horizontal dotted line in the top panel). With VRR included,
  high-eccentricity oscillations occur as a consequence of LK-like evolution
  induced by high mutual inclinations triggered by VRR\@. In particular, a flip
  from prograde to retrograde orientation occurs after
  $\simeq 85\,t_\mathrm{LK}$. This flip is associated with a maximum
  eccentricity of $\simeq 0.99$, much larger than $\simeq 0.84$ in the case
  without VRR\@. }
\label{fig:example_0}
\end{figure}

\begin{figure}
\center
\includegraphics[width=0.52\textwidth]{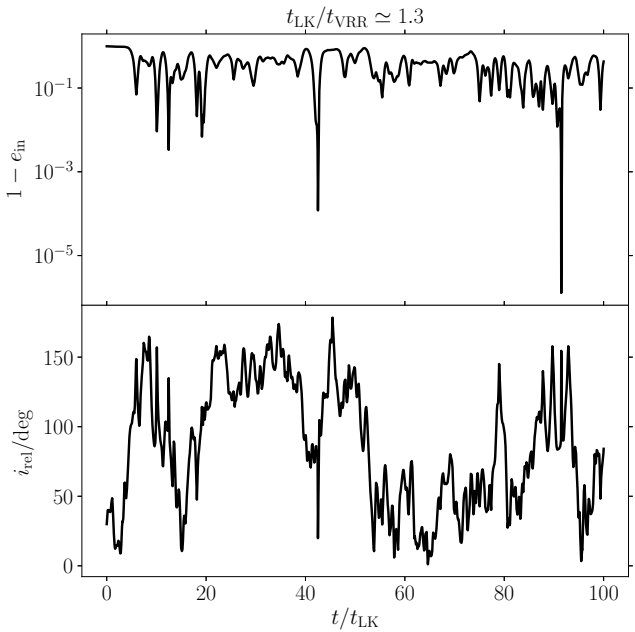}
\caption{\small Example evolution in the `transadiabatic' regime,
  $\mathcal{R}\simeq 1.3$. Top panel: inner orbit eccentricity; bottom panel:
  mutual inclination between the inner and outer orbits. The initial mutual
  inclination is $i_\mathrm{rel,\,0} \simeq 30^\circ$, implying no excitation
  in the case without VRR, yet with VRR, high-eccentricity oscillations occur,
  reaching values as high as $1-10^{-6}$. }
\label{fig:example_1}
\end{figure}

We first show a number of examples to illustrate the dependence of the typical
dynamical behavior on the adiabatic parameter $\mathcal{R}$. In these examples, we fix
$e_\oout = 2/3$. In the limit that
$\mathcal{R}$ is very small, i.e., $t_\mathrm{LK} \ll t_\mathrm{VRR}$, VRR is
unimportant, and, to quadrupole order in the test particle approximation, the
maximum eccentricity, assuming zero initial eccentricity, is given by the canonical relation
\begin{align}
\label{eq:e_max_LK}
  e_\mathrm{in,\max} = \sqrt{1-\frac{5}{3} \cos^2(i_\mathrm{rel,\,0})},
\end{align}
where $i_\mathrm{rel,\,0}$ is the initial mutual inclination. If $\mathcal{R}$
is larger, in particular, $\mathcal{R} \sim 1$, significant eccentricity
excitation is possible even if the initial mutual inclination is small,
including coplanar configurations. This is illustrated in
Figures~\ref{fig:example_0} and~\ref{fig:example_1}, in which
$\adpar \simeq 0.11$ and $\simeq 1.3$, respectively. In these examples, there
is significant eccentricity excitation, much more than implied by
equation~(\ref{eq:e_max_LK}). In particular, in \F~\ref{fig:example_1}, the
initial mutual inclination is $i_\mathrm{rel,\,0} \approx 30^\circ$, and there
would be no eccentricity excitation in the absence of VRR\@. If VRR is taken
into account (with $\mathcal{R} \simeq 1.3$), the mutual inclination changes
due to the changing direction of $\ve{j}_\oout$. After a few $t_\mathrm{LK}$,
$i_\mathrm{rel}$ increases to $90^\circ$, and the eccentricity is excited
through a LK-like process. Subsequent eccentricity excitations are associated
with inclinations close to $90^\circ$, and particularly with orbital flips
(changes from prograde to retrograde orbits, and vice versa). For example, a
spike in the eccentricity with $1-e_\oin$ reaching $\sim 10^{-6}$ occurs after
$\approx 90\, t_\mathrm{LK}$, and which is associated with an orbital
flip. Typically, the evolution is complex and chaotic in this regime. High
eccentricities are possible even for $\adpar \simeq 0.1$, as shown in
\F~\ref{fig:example_0}.

\begin{figure}
  \center
  \includegraphics[width=0.52\textwidth]{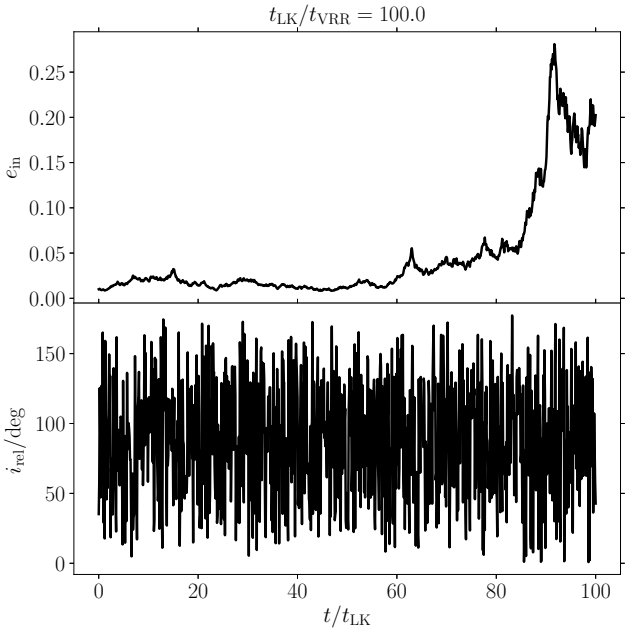}
  \caption{\small Similar to \F~\ref{fig:example_1}, now showing example
    evolution in the regime $\mathcal{R} \gg 1$ ($\mathcal{R}=100$). Initially,
    eccentricity excitation is suppressed, but the eccentricity growths slowly
    in a diffusive process.  }
\label{fig:example_2}
\end{figure}

In \F~\ref{fig:example_2}, we illustrate the regime $\mathcal{R} \gg 1$
($\mathcal{R}=100$). One might expect that eccentricity excitation is
completely suppressed if $t_\mathrm{LK} \gg t_\mathrm{VRR}$, since the mutual
inclination is varying on a short time-scale compared to $t_\mathrm{LK}$ (see
the bottom panel in \F~\ref{fig:example_1}). Initially, this is the case in
\F~\ref{fig:example_2}, but after several tens of $t_\mathrm{LK}$, the
eccentricity grows gradually. The evolution in this regime can be characterized
as a diffusive process, in which the eccentricity grows in a random walk-like
fashion with a step size given by
$\Delta e_\mathrm{step} \sim t_\mathrm{VRR}/t_\mathrm{LK} = \adpar^{-1}$. The
eccentricity then grows as
$\Delta e = \Delta e_\mathrm{step} {(\Delta t/t_\mathrm{VRR})}^{1/2}$, implying
that $e$ increases to order unity on a time-scale given by (setting
$\Delta e = 1$ for $\Delta t = t_\mathrm{dif}$)
\begin{align}
t_\mathrm{dif} \sim \adpar^2\,t_\mathrm{VRR} = \adpar \, t_\mathrm{LK}.
\end{align}
In \F~\ref{fig:example_2}, $\adpar=100$, and the eccentricity indeed grows to
order unity on a time-scale of the order of
$\adpar \, t_\mathrm{LK} = 100\,t_\mathrm{LK}$.

\subsection{Eccentricity distributions as a function of the adiabatic parameter}
\label{sect:dyn:reg}
We investigate the different regimes more quantitatively by carrying out
numerical integrations of the equations of motion for different adiabatic
parameters, again only including the secular Newtonian three-body terms and
VRR\@. Specifically, for a given value of the adiabatic parameter $\adpar$ in a
range of up to $\adpar=100$, we sample 1000 systems with different random
orientations of the inner binary with respect to the MBH, and integrate for a
duration of $100\,t_\mathrm{LK}$. Note that, by definition, an elapsed time of
$t_\mathrm{LK}$ corresponds to $\adpar \,t_\mathrm{VRR}$, i.e., in the longest
integration ($\adpar=100$) we integrate for $10^4\,t_\mathrm{VRR}$. For each
system, we record the maximum eccentricity reached in the inner orbit,
$e_{\mathrm{in},\,\max}$, by means of root finding (i.e., we search for all local
maxima in the integrations, $\mathrm{d}e_\mathrm{in}/\mathrm{d}t=0$, using root
finding, and determine from the local maxima the global maximum; this method
ensures that there is no risk of missing the true maximum eccentricity because
of a finite number of output snapshots). For simplicity, we fix $e_\oout=2/3$.

\begin{figure}
\center
\includegraphics[width=0.52\textwidth]{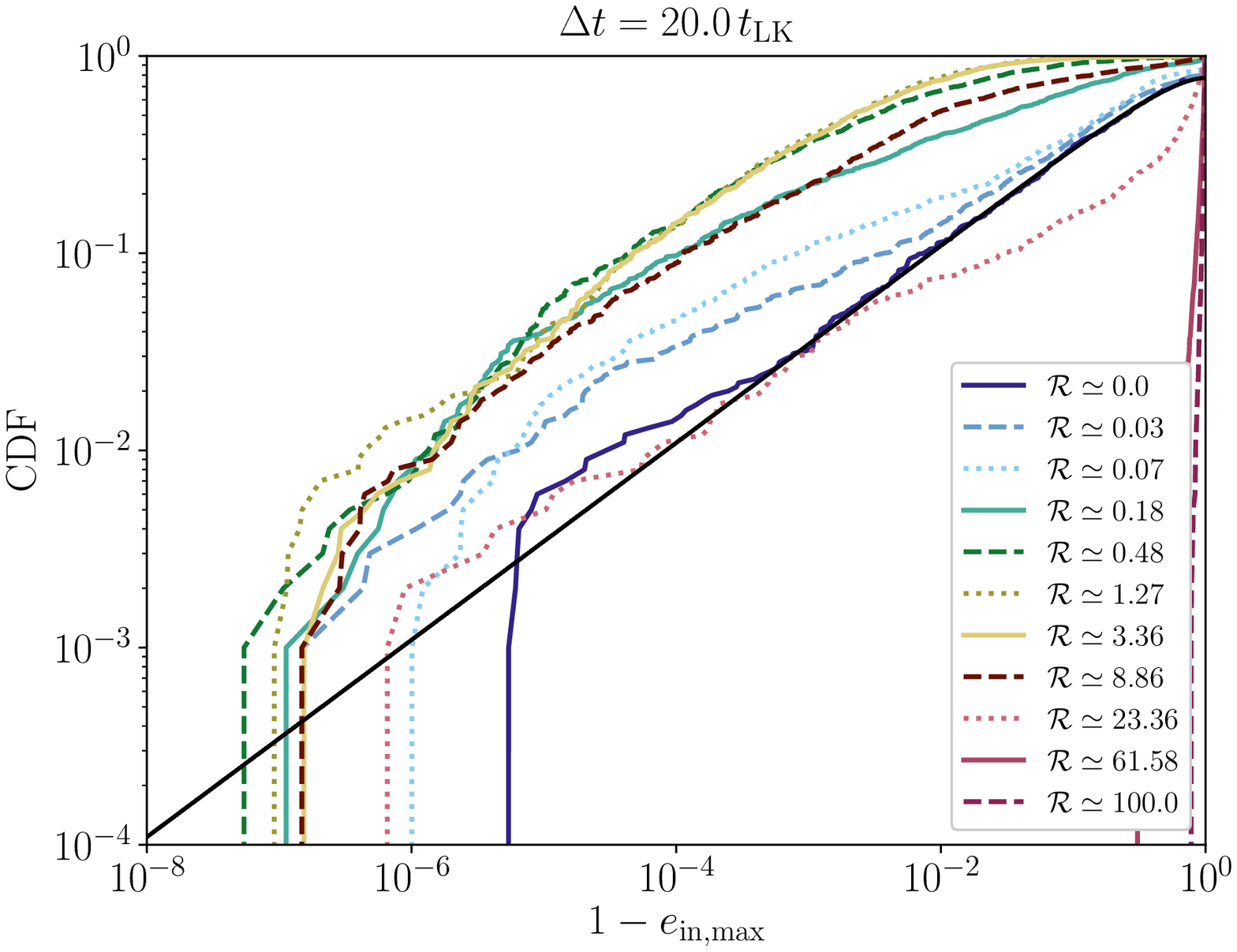}
\includegraphics[width=0.52\textwidth]{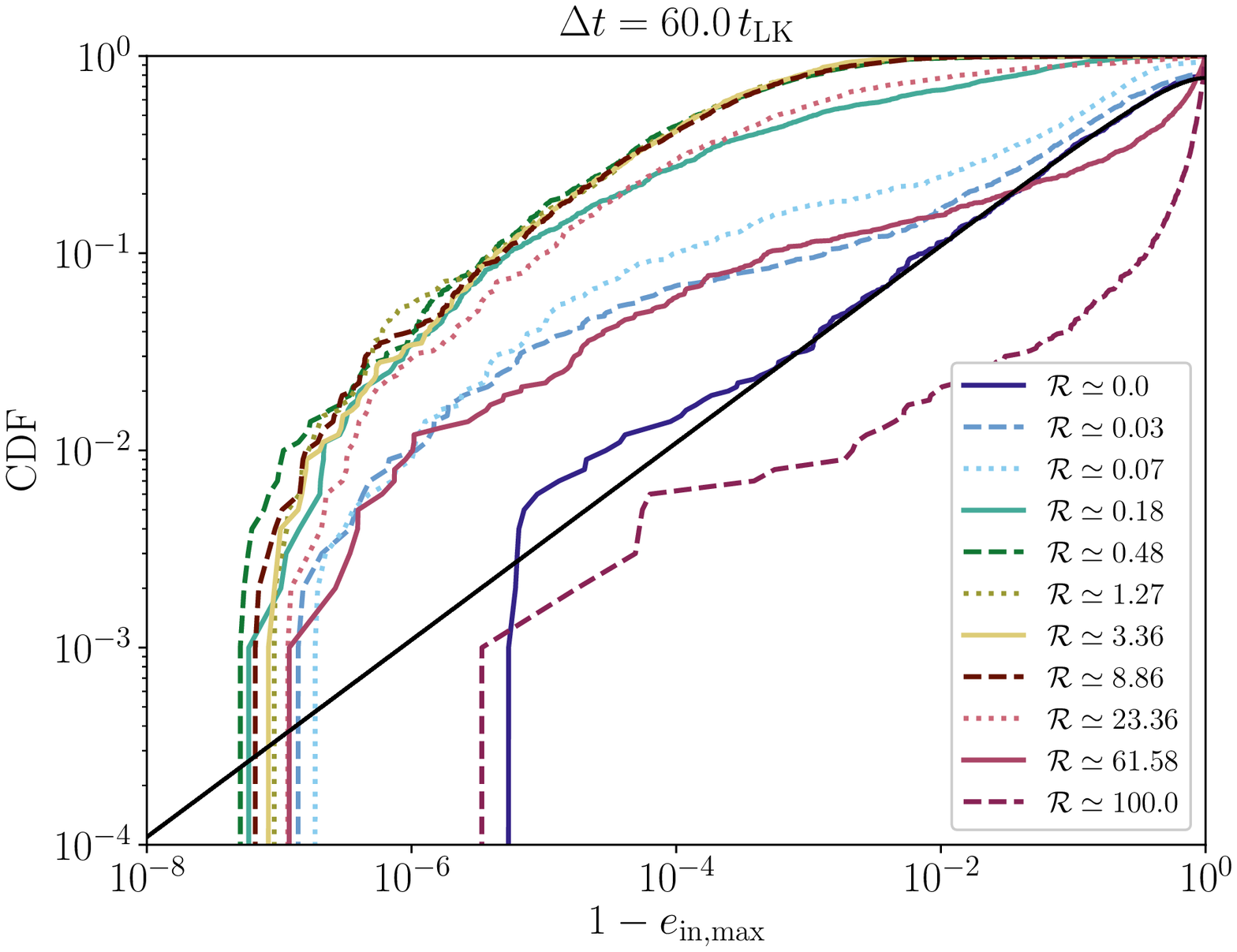}
\includegraphics[width=0.52\textwidth]{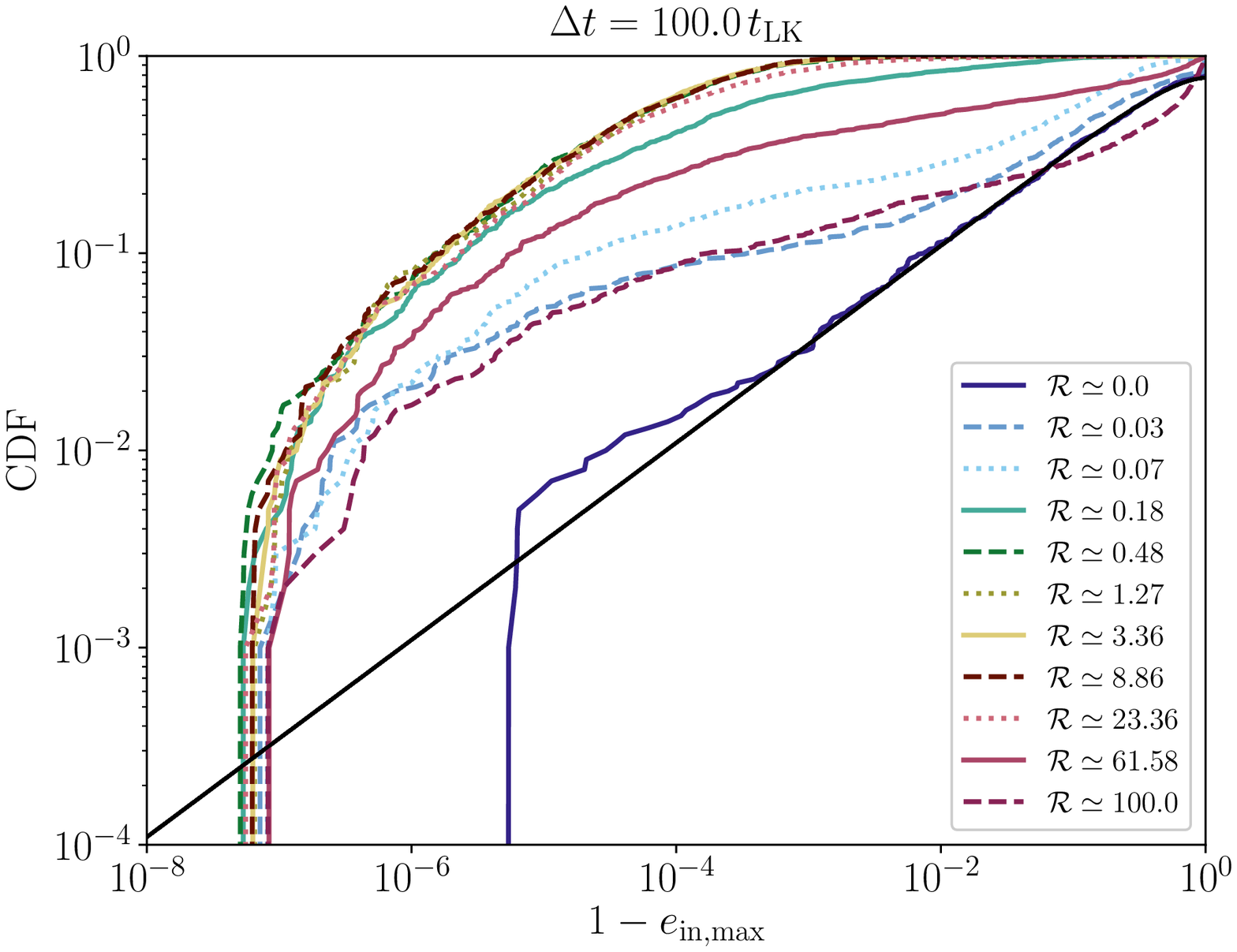}
\caption{\small Cumulative distributions of $e_{\mathrm{in},\,\max}$ (based on
  1000 simulations). Each panel corresponds to a different integration time,
  indicated in the top. Different lines correspond to different values of
  $\adpar$, indicated in the legend. Black solid line: analytic distribution
  that applies in the limit $\adpar=0$ (see equation~\ref{eq:cdf_emax_R0}). }
\label{fig:emax_cdfs}
\end{figure}

The resulting cumulative distributions of $e_{\mathrm{in},\,\max}$ (each based on
1000 simulations) are are shown in \F~\ref{fig:emax_cdfs}. Each panel
corresponds to a different integration time, indicated in the top. Different
lines correspond to different values of $\adpar$, indicated in the legend. The
black solid line shows the analytic result that applies in the limit $\adpar=0$
(i.e., the canonical LK limit without VRR). In that case, it is straightforward
to show that
\begin{align}
\label{eq:cdf_emax_R0}
N(<j_\mathrm{in}) = \sqrt{3/5} \, j_\mathrm{in},
\end{align}
where $j_\mathrm{in} = \sqrt{1-e_\mathrm{in}^2}$. The solid dark blue lines in
\F~\ref{fig:emax_cdfs} corresponds to the numerical integrations with
$\adpar=0$, and agree with the analytical prediction (note that the noise level
is at a CDF value of $\sim 1/\sqrt{1000} \simeq 0.03$). We identify the
following features.

\begin{itemize}
\item Already for relatively small $\adpar$ ($\adpar \simeq 0.03$), there is a
  significant enhancement in the maximum eccentricity compared to
  equation~(\ref{eq:cdf_emax_R0}). For example, after $100\,t_\mathrm{LK}$ and
  for $1-e_{\mathrm{in},\,\max}= 10^{-4}$, the cumulative fraction is
  $\sim 10^{-2}$ in the case $\adpar=0$, whereas for $\adpar \simeq 0.03$, the
  cumulative fraction at $1-e_{\mathrm{in},\,\max}= 10^{-4}$ is $\sim 10^{-1}$.
\item Generally, there is a dependence of the distribution of
  $e_{\mathrm{in},\,\max}$ on time. As $\adpar$ is closer to unity, the dependence
  becomes weaker, i.e., there is less evolution in the distributions between 20
  and 100 $t_\mathrm{LK}$.
\item For a range of $\adpar$, $0.1\lesssim \adpar \lesssim 10$, the
  distribution of $e_{\mathrm{in},\,\max}$ is approximately independent of
  $\adpar$. Related to the above point, the time evolution of this distribution
  is relatively weak compared to other values of $\adpar$. After
  $100\,t_\mathrm{LK}$, at $1-e_{\mathrm{in},\,\max}= 10^{-4}$ the cumulative
  fraction for these values of $\adpar$ is $\sim 0.6$, compared to
  $\sim 10^{-2}$ for the case without VRR\@.
\item For large values of $\adpar$, $\adpar \gtrsim 10$, there is a suppression
  of maximum eccentricities compared to equation~(\ref{eq:cdf_emax_R0}) at
  early times. At later times, the distributions shift significantly to larger
  eccentricities. After $100\,t_\mathrm{LK}$ and for $\adpar=100$, the median
  $e_{\mathrm{in},\,\max}$ is comparable to the case $\adpar=0$, but there is a
  tail towards very high eccentricities which extends to about one order of
  magnitude in the cumulative fraction at
  $1-e_{\mathrm{in},\,\max} \sim 10^{-5}$. This can be ascribed to the
  diffusive evolution mentioned previously in \S~\ref{sect:dyn:ex}.
\end{itemize}

\section{Monte Carlo simulations}
\label{sect:pop_syn}
In addition to the scale-free integrations of \S~\ref{sect:dyn}, we carry out
Monte Carlo simulations of BH-BH binaries around MBHs in which we include
more (astro)physical effects. In particular, we include PN terms (1PN terms in
the inner and outer orbits, and 2.5PN and spin-orbit coupling terms in the
inner orbit), Newtonian mass precession terms for the outer orbit, and take
into account binary evaporation (with a simplified method, i.e., by limiting the
integration time).

\subsection{Setup}
\label{sect:pop_syn:setup}
\subsubsection{Initial conditions}
\label{sect:pop_syn:setup:IC}
We carry out various sets of simulations, each in which we consider the MBH
mass and density slope $\gamma$ to be fixed parameters. For a given $M_\bullet$
and $\gamma$, we first calculate the radius of the sphere of influence,
$r_\mathrm{h}$, using the same method of \S~\ref{sect:time_scales:comp}, i.e.,
we assume a stellar density distribution
$n_\star(r) = n_\mathrm{h} {(r/r_\mathrm{h})}^{-\gamma}$, and normalize the
distribution using the $M_\bullet$--$\sigma_\mathrm{bulge}$-relation.

The initial distributions of compact object binaries in galactic nuclei are
not constrained. Here, we assume simple distributions that do not impose
strong biases on the parameters. We emphasize that our assumptions on the
distributions are not realistic, and are chosen for simplicity and generality,
rather than for realism. A posteriori, our merger rates can be scaled accordingly
 to more specific distributions. 

We generate $N_\mathrm{MC}$ systems with the following assumptions. First, we
sample the primary mass $M_1$ from a flat distribution with $10<M_1/\msun<50$,
approximately capturing the low-mass range of BHs, and also including the
possibility of more massive BHs (the highest BH mass detected by LIGO to date
is $\approx 35 \,\msun$ for GW150914, \citealt{2016PhRvL.116f1102A}). We note that
we find few mergers with primary masses lower than 10 and higher than 50 $\msun$. The
secondary mass is sampled assuming a flat distribution of the mass ratio
$q\equiv M_2/M_1$, with a minimum secondary mass of $M_2=10\,\msun$. A flat
mass ratio distribution is appropriate for massive (main-sequence) stars
\citep{2012Sci...337..444S,2013ARAA..51..269D,2014ApJS..213...34K}, although it likely
does not apply to compact object binaries. We note, however, that there is no discernible 
preference for mergers regarding $q$, as shown below.

The inner and outer semimajor axes are sampled from a flat distribution in
$\log_{10}(a_i/\mathrm{AU})$ (i.e., \"{O}pik's law,
\citealt{1924PTarO..25f...1O}). For the inner orbits, we set the range to
$1<a_\oin/\au<10^4$. The lower limit is motivated from the result of
\S~\ref{sect:time_scales} that the number of potentially interesting systems
drops rapidly as $a_\oin \lesssim 10\,\au$ due to general relativistic
precession. This is also supported a posteriori from the simulations, given
that we find few mergers with initial inner orbit semimajor axes less than a
few $\au$. The upper limit is motivated by the fact that such very wide
binaries quickly evaporate, and/or are dynamically unstable with respect to the MBH.
The range of $a_\oout$ is set to
$0.01<a_\oout/r_\mathrm{h}<2$. For smaller $a_\oout$, we find few
mergers a posteriori, and our models do not apply for $a_\oout \gg r_\mathrm{h}$.

We note that our distribution of the semimajor axis of the compact object binary
differs from simulations of globular clusters, which typically predict distributions
that are peaked around $\sim 0.1$ to $\sim 1 \,\au$ (for binaries ejected from the cluster,
e.g., \citealt{2016PhRvD..93h4029R}). In the case of field binaries, the distribution is also
biased towards small $a_\oin$, given that wider binaries are more susceptible to becoming
unbound due to supernovae. This implies that the merger fractions would be lower if we had
adopted such distributions for the initial semimajor axis of the compact object binary. This is borne
out by a set of exploratory simulations that we carried out in which the inner binary was evolved from the
MS until a compact object binary was formed using the \textsc{BSE} binary population synthesis code
\citep{2002MNRAS.329..897H}, similar to the approach of \citet{2017ApJ...846..146P}. In this exploratory
set of simulations, the merger fractions were lower by a factor of $\sim 10$, and this resulted in poor
number statistics for the merging systems. 

The inner and outer orbit eccentricities are sampled from
a thermal distribution, $\mathrm{d} N/\mathrm{d} e_i=2e_i$ (with
$0.01<e_i<0.95$). A thermal distribution is to be expected for the outer orbit due to dynamical
interactions with the stellar cluster \citep{1919MNRAS..79..408J}. Regarding the inner
orbit, we expect that a significant eccentricity would typically be produced in the relatively wide
($a_\oin>1\,\au$) orbit due to the supernova kick associated with the formation of the
second compact object. We note that there is little dependence
of our results on the initial eccentricities. 

Sampled systems that are dynamically unstable are rejected, where we use the
stability criterion of \citet{2001MNRAS.321..398M}. In this approach, none of
the systems are initially tidally disrupted by the MBH, i.e., in all cases
$a_\oout (1-e_\oout) > a_\oin {[M_\bullet/(M_1+M_2)]}^{1/3}$. We compute the
evaporation time-scale for the binary using equation~(\ref{eq:t_EV}), where we
set $\sigma_\star = \sigma_\star(r_\mathrm{h})$ if $r_\oout>r_\mathrm{h}$ to
take into account that the velocity dispersion outside of the radius of
influence is typically constant.

The spins of the two compact objects are assumed to be initially coplanar with
the inner orbit. This is likely not a good assumption if natal kicks were involved in
the formation of the compact object binary. However, our aim is to show that LK-VRR
coupling can very efficiently induce random spin-orbit orientations (see
 \S\,\ref{sect:pop_syn:results:orbits} below). Therefore, we choose to set up the system with
  zero initial spin-orbit alignments. Any randomization in the spin-orbit alignment can
   then clearly be attributed to the dynamical
evolution. The spins are evolved according to the orbit-averaged PN
geodetic equations of motion (e.g., \citealt{2004PhRvD..70l4020S}), where we
neglect the back-reaction of the spins on the orbit.

Each simulation is carried out twice: once with VRR included, and once without
the effects of VRR\@. We consider two sets of simulations. In the `small' set,
we set the number of BH-BH binaries to be $N_\mathrm{MC}=10^3$,
 and consider 20 different combinations of
$M_\bullet$ and $\gamma$. The number of systems for each parameter combination
in this set is relatively low, implying low merger number statistics. We also
consider a `large' set, in which we set $N_\mathrm{MC}=10^4$. In the large set,
we consider the combination $M_\bullet=10^4\,\msun$ and $\gamma=2$ (to maximize
the absolute number of mergers), and the combination
$M_\bullet = 4 \times 10^6\,\msun$ and $\gamma=1.3$, representative of the
Galactic Center (GC\@;
\citealt{2008ApJ...689.1044G,2009ApJ...692.1075G,2016ApJ...821...44F,2018AA...609A..27S}).

\subsubsection{Stopping conditions}
\label{sect:pop_syn:setup:sc}
The successfully-sampled systems are integrated for a duration of
$10\,\mathrm{Gyr}$ or the evaporation time-scale (equation~\ref{eq:t_EV}),
whichever is shortest. In addition, we check for the following conditions
during the simulations using root finding within the numerical integration of
the ODEs (this ensures that no conditions are missed due to finite output
times).
\begin{enumerate}
\item The time-scale for GW emission in the inner binary to shrink $a_\oin$ by
  order itself, $t_{a,\,\mathrm{GW}}$, is 10 times shorter than the time-scale
  for LK oscillations to reduce the inner binary periapsis distance
  $r_{\mathrm{p},\,\oin}$ by order itself,
  $t_{r_\mathrm{p},\,\mathrm{LK}}$. Specifically,
  $10\,t_{a,\,\mathrm{GW}} < t_{r_\mathrm{p},\,\mathrm{LK}}$, where
\begin{align}
  \nonumber \label{eq:t_a_GW}t_{a,\,\mathrm{GW}}^{-1} &\equiv \left | \frac{1}{a_\oin} \frac{\mathrm{d} a_\oin}{\mathrm{d} t} \right |_\mathrm{GW} \\
                                                      &= \frac{64}{5} \frac{G^3 M_1 M_2(M_1+M_2) }{c^5 a_\oin^4 {\left(1-e_\oin^2 \right )}^{7/2}} \left (1 + \frac{73}{24} e_\oin^2 + \frac{37}{96} e_\oin^4 \right ); \\
  \nonumber \label{eq:t_rp_LK} t_{r_\mathrm{p},\,\mathrm{LK}}^{-1} &\equiv \left | \frac{1}{r_{\mathrm{p},\,\oin}} \frac{\mathrm{d} r_{\mathrm{p},\,\oin}}{\mathrm{d} t} \right |_\mathrm{LK} \\
  \nonumber &= \frac{75}{64} \sqrt{\frac{5}{3}} \frac{e_\oin}{\sqrt{1-e_\oin^2}} \sqrt{\frac{G (M_1+M_2)}{a_\oin^3}} \frac{M_\bullet}{M_1+M_2} {\left ( \frac{a_\oin}{a_\oout} \right )}^3 \\
                                                      &\quad \times {\left (1-e_\oout^2 \right )}^{-3/2}.
\end{align}
Equation~(\ref{eq:t_a_GW}) follows directly from equation~(5.6) of
\citet{1964PhRv..136.1224P}; to derive equation~(\ref{eq:t_rp_LK}), we computed
the rms average of the secular quadrupole-order LK equation for $\dot{e}_\oin$
over all directions of $\unit{J}_\oout$.

Once this condition is met, the binary is effectively decoupled from the MBH,
and the evolution is dominated by the 2.5PN GW terms in the inner binary. We
stop the subsequent integration, and flag the system as a compact object binary
merger. The factor 10 in
$10\,t_{a,\,\mathrm{GW}} < t_{r_\mathrm{p},\,\mathrm{LK}}$ is a `safety' factor
to ensure that the system is completely decoupled. We demonstrate that our
criterion is robust by giving an example below in
\S~\ref{sect:pop_syn:example}.
\item The secular approximation in the orbit-averaged three-body equations of
  motion breaks down (i.e., the semisecular or quasi-secular regime is
  reached). This can occur when the time-scale for the inner orbit
  angular-momentum to change by order itself becomes comparable to or even
  shorter than the inner or outer orbital periods, in which case averaging over
  the orbits is clearly no longer a good approximation. We assume that the
  semisecular regime is entered if \citep{2014ApJ...781...45A}
\begin{align}
\label{eq:semisec}
  \sqrt{1-e_\oin} < 5 \pi \frac{M_\bullet}{M_1+M_2} {\left [ \frac{a_\oin}{a_\oout \left(1-e_\oout \right )} \right ]}^3.
\end{align}
In the semisecular regime, the system can remain dynamically stable, but the
maximum eccentricities reached in the inner orbit can be potentially higher
compared to the maximum eccentricities according to the doubly-averaged
equations of motion \citep{2012ApJ...757...27A,2014ApJ...781...45A}. Here, we continue integrating
the equations of motion using the secular equations, even after the semisecular
regime is reached. We do record when the system enters the semisecular
regime. This implies that our merger rates are likely underestimated. However,
our focus is on the relative enhancement of the rates when VRR is taken into
account. We do not expect that this approach introduces significant errors in
the relative merger rates.
\end{enumerate}

Also, we stop the integration if the system becomes dynamically unstable
according to the criterion of~\citet{2001MNRAS.321..398M}. Although we do check
for this condition, it does not occur in the simulations since
originally-unstable systems were rejected, and $a_\oout$, $e_\oout$ and the
masses are assumed to be constant (also, note that $a_\oin$ can only decrease
due to the 2.5PN terms, bringing the system even further away from dynamical
instability). Similarly, we check for tidal disruption of the binary by the
MBH, i.e., $a_\oout (1-e_\oout) < a_\oin {[M_\bullet/(M_1+M_2)]}^{1/3}$, which
does not occur in the simulations for similar reasons.

\subsection{Results}
\label{sect:pop_syn:results}

\subsubsection{Example}
\label{sect:pop_syn:example}

\begin{figure}
\center
\includegraphics[width=0.52\textwidth]{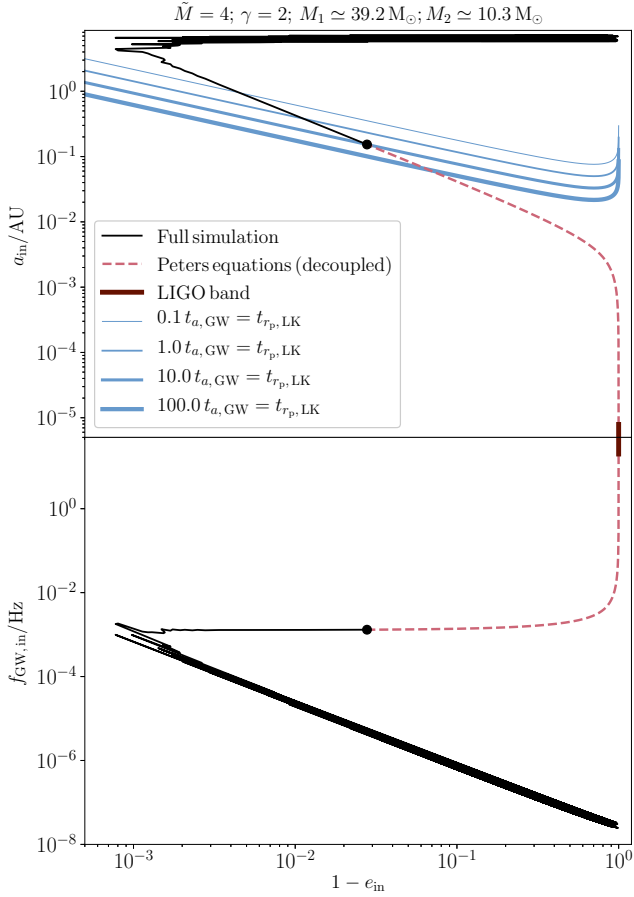}
\caption{\small Example evolution of a system in which the inner binary merges
  due to coupled LK-VRR evolution. The top and bottom panels show evolutionary
  tracks in the $(1-e_\oin,a_\oin)$ and $(1-e_\oin,f_\mathrm{GW,\, \oin})$
  planes, respectively, where $f_\mathrm{GW,\,\oin}$ is the inner orbit GW
  frequency (see equation~\ref{eq:f_GW}). Refer to the text for the initial
  conditions. Solid black lines: evolution according to the `full' numerical
  simulations (including the three-body secular terms, terms associated with
  VRR, and PN terms); the black circles correspond to stopping condition (1) in
  \S~\ref{sect:pop_syn:setup:sc}. Red dashed lines: analytic solutions for
  2.5PN evolution in isolated binaries according to
  \citet{1964PhRv..136.1224P}. The thick red solid lines indicate the LIGO
  band ($f_\mathrm{GW,\,\oin}>20\,\mathrm{Hz}$). In the top panel, the blue
  solid curves show contours for which $t_{a,\,\mathrm{GW}}$
  (equation~\ref{eq:t_a_GW}) and $t_{r\mathrm{p},\,\mathrm{LK}}$
  (equation~\ref{eq:t_rp_LK}) are related by various factors, indicated in the
  legend. }
\label{fig:pop_syn_example}
\end{figure}

In \F~\ref{fig:pop_syn_example}, we give an example of the evolution in the
$(1-e_\oin,a_\oin)$ plane (top panel), and the
$(1-e_\oin,f_\mathrm{GW,\, \oin})$ plane (bottom panel), of a system in which
the inner binary eventually merges. Here, $f_\mathrm{GW,\, \oin}$ is the peak
GW frequency in the inner binary (i.e., the mode with the most power),
calculated according to \citet{2003ApJ...598..419W}
\begin{align}
\label{eq:f_GW}
f_\mathrm{GW,\, \oin} =  \frac{1}{\pi} \sqrt{ \frac{ G(M_1+M_2)}{a_\oin^3}} \frac{ {(1+e_\oin)}^{1.1954}}{ {\left(1-e_\oin^2 \right )}^{3/2}}.
\end{align}
The initial parameters in the example system are $M_\bullet=10^4\,\msun$,
$\gamma=2$, $M_1 \simeq 39.2 \, \msun$, $M_2 \simeq 10.3\,\msun$,
$a_\oin \simeq 7.13\,\au$,
$a_\oout \simeq 2.5 \times 10^3 \, \au \simeq 0.012 \, \mathrm{pc}$,
$e_\oin \simeq 0.15$, $e_\oout \simeq 0.43$, and
$i_\mathrm{rel} \simeq 80^\circ$. The LK and VRR time-scales are comparable
($t_\mathrm{LK} \simeq 0.068 \, \mathrm{Myr}$;
$t_\mathrm{VRR} \simeq 0.11\, \mathrm{Myr}$, such that $\adpar \simeq 0.63$);
in the example, VRR is included, and high eccentricities are reached due to
coupled LK-VRR evolution. Note that, although the initial relative inclination
is relatively high, it is not high enough to trigger a merger in the absence of
VRR\@.

As shown in \F~\ref{fig:pop_syn_example}, the inner binary semimajor axis
decreases slightly due to the 2.5PN terms each time the eccentricity is high,
resulting in a wide horizontal band in the $(1-e_\oin,a_\oin)$ plane. After
$\sim 680 \, \mathrm{Myr}$ of evolution, the 2.5PN terms become more dominant
and the binary starts to spiral in due to GW emission. The eccentricity
oscillations then stop as $a_\oin$ decreases, which can be ascribed to the
increasing secular time-scale $t_\mathrm{LK}$ with decreasing $a_\oin$, whereas
the 1PN precession time-scale decreases.

The end of the `full' simulation, i.e., in which the three-body secular terms,
terms associated with VRR, and PN terms are included, is marked with the black
circles in \F~\ref{fig:pop_syn_example}. At this stage, the inner binary is
clearly decoupled from perturbations of the MBH and the surrounding stellar
cluster, and the evolution is dominated by the 2.5PN terms. The subsequent
evolution is not computed in the `full' simulation, but shown in
\F~\ref{fig:pop_syn_example} with the red dashed lines, which are based on the
analytic solution of \citet{1964PhRv..136.1224P}, i.e.,
\begin{align}
\label{eq:ae_peters}
  a_\oin(e_\oin) = \frac{c_0 e_\oin^{12/19}}{1-e_\oin^2} {\left ( 1 + \frac{121}{304} e_\oin^2 \right )}^{870/2299},
\end{align}
where $c_0$ follows from the initial conditions (i.e., the black circles in
\F~\ref{fig:pop_syn_example}). The parts of the curves corresponding to the
LIGO band (i.e., a GW frequency higher than 20 Hz) are shown with thick solid
dark red lines. The eccentricity when reaching the LIGO band,
$e_{\oin,\,\mathrm{LIGO}} \approx 2 \times 10^{-5}$, is small; this is
generally the case in the Monte Carlo simulations (see
\S~\ref{sect:pop_syn:results:orbits} below).

In the top panel of \F~\ref{fig:pop_syn_example}, we also show with four solid
blue lines several contours for which the GW and LK time-scales are related by
factors ranging between 0.1 and 100 (refer to the legend). Since the `full'
simulations were stopped when
$10 \, t_{a,\,\mathrm{GW}} = t_{r_\mathrm{p},\,\mathrm{LK}}$, the third contour
intersects with the black circle. The contours show that there is not a strong
sensitivity of the assumed factor between the time-scales, on whether or not
the system would have been identified as a merger system. For example, if the
criterion $t_{a,\,\mathrm{GW}} = t_{r_\mathrm{p},\,\mathrm{LK}}$ had been
adopted, the `full' simulation would have been stopped at a larger $a_\oin$,
but still at a point in the $(1-e_\oin,a_\oin)$-space in which the inner binary
is decoupled.

\subsubsection{Fractions}
\label{sect:pop_syn:results:fractions}

\definecolor{Gray}{gray}{0.9}
\begin{table*}
\begin{tabular}{lcccccccc}
  \toprule
  & \multicolumn{8}{c}{Fraction} \\
  & \multicolumn{2}{c}{No Interaction} & \multicolumn{2}{c}{No Interaction} & \multicolumn{2}{c}{Merger} & \multicolumn{2}{c}{Merger} \\
  & & & \multicolumn{2}{c}{Semisecular} & & & \multicolumn{2}{c}{Semisecular} \\
  Fixed parameters & VRR & No VRR & VRR & No VRR & VRR & No VRR & VRR & No VRR \\
  \midrule
  $N_\mathrm{MC}=10^3$ \\
  \midrule
  $\tilde{M}_\bullet = 4.0; \, \gamma = 1.3$ & $0.817 \pm 0.029$ & $0.976 \pm 0.031$ & $0.018 \pm 0.004$ & $0.006 \pm 0.002$ & $0.132 \pm 0.011$ & $0.013 \pm 0.004$ & $0.033 \pm 0.006$ & $0.005 \pm 0.002$ \\
  $\tilde{M}_\bullet = 4.0; \, \gamma = 2.0$ & $0.680 \pm 0.026$ & $0.965 \pm 0.031$ & $0.034 \pm 0.006$ & $0.012 \pm 0.003$ & $0.224 \pm 0.015$ & $0.019 \pm 0.004$ & $0.062 \pm 0.008$ & $0.004 \pm 0.002$ \\
  $\tilde{M}_\bullet = 4.3; \, \gamma = 1.3$ & $0.904 \pm 0.030$ & $0.986 \pm 0.031$ & $0.014 \pm 0.004$ & $0.004 \pm 0.002$ & $0.063 \pm 0.008$ & $0.007 \pm 0.003$ & $0.019 \pm 0.004$ & $0.003 \pm 0.002$ \\
  $\tilde{M}_\bullet = 4.3; \, \gamma = 2.0$ & $0.811 \pm 0.029$ & $0.974 \pm 0.031$ & $0.024 \pm 0.005$ & $0.015 \pm 0.004$ & $0.136 \pm 0.012$ & $0.009 \pm 0.003$ & $0.029 \pm 0.005$ & $0.002 \pm 0.001$ \\
  $\tilde{M}_\bullet = 4.7; \, \gamma = 1.3$ & $0.952 \pm 0.031$ & $0.988 \pm 0.031$ & $0.011 \pm 0.003$ & $0.004 \pm 0.002$ & $0.030 \pm 0.005$ & $0.006 \pm 0.002$ & $0.007 \pm 0.003$ & $0.002 \pm 0.001$ \\
  $\tilde{M}_\bullet = 4.7; \, \gamma = 2.0$ & $0.907 \pm 0.030$ & $0.988 \pm 0.031$ & $0.019 \pm 0.004$ & $0.007 \pm 0.003$ & $0.048 \pm 0.007$ & $0.004 \pm 0.002$ & $0.026 \pm 0.005$ & $0.001 \pm 0.001$ \\
  $\tilde{M}_\bullet = 5.0; \, \gamma = 1.3$ & $0.958 \pm 0.031$ & $0.988 \pm 0.031$ & $0.015 \pm 0.004$ & $0.008 \pm 0.003$ & $0.021 \pm 0.005$ & $0.004 \pm 0.002$ & $0.006 \pm 0.002$ & $0.000 \pm 0.000$ \\
  $\tilde{M}_\bullet = 5.0; \, \gamma = 2.0$ & $0.947 \pm 0.031$ & $0.990 \pm 0.031$ & $0.012 \pm 0.003$ & $0.007 \pm 0.003$ & $0.026 \pm 0.005$ & $0.002 \pm 0.001$ & $0.015 \pm 0.004$ & $0.001 \pm 0.001$ \\
  $\tilde{M}_\bullet = 5.3; \, \gamma = 1.3$ & $0.982 \pm 0.031$ & $0.996 \pm 0.032$ & $0.006 \pm 0.002$ & $0.002 \pm 0.001$ & $0.009 \pm 0.003$ & $0.002 \pm 0.001$ & $0.003 \pm 0.002$ & $0.000 \pm 0.000$ \\
  $\tilde{M}_\bullet = 5.3; \, \gamma = 2.0$ & $0.972 \pm 0.031$ & $0.994 \pm 0.032$ & $0.006 \pm 0.002$ & $0.002 \pm 0.001$ & $0.014 \pm 0.004$ & $0.002 \pm 0.001$ & $0.008 \pm 0.003$ & $0.002 \pm 0.001$ \\
  $\tilde{M}_\bullet = 5.7; \, \gamma = 1.3$ & $0.991 \pm 0.031$ & $0.997 \pm 0.032$ & $0.003 \pm 0.002$ & $0.001 \pm 0.001$ & $0.005 \pm 0.002$ & $0.002 \pm 0.001$ & $0.001 \pm 0.001$ & $0.000 \pm 0.000$ \\
  $\tilde{M}_\bullet = 5.7; \, \gamma = 2.0$ & $0.981 \pm 0.031$ & $0.993 \pm 0.032$ & $0.008 \pm 0.003$ & $0.003 \pm 0.002$ & $0.008 \pm 0.003$ & $0.002 \pm 0.001$ & $0.003 \pm 0.002$ & $0.002 \pm 0.001$ \\
  $\tilde{M}_\bullet = 6.0; \, \gamma = 1.3$ & $0.988 \pm 0.031$ & $0.994 \pm 0.032$ & $0.005 \pm 0.002$ & $0.002 \pm 0.001$ & $0.007 \pm 0.003$ & $0.004 \pm 0.002$ & $0.000 \pm 0.000$ & $0.000 \pm 0.000$ \\
  $\tilde{M}_\bullet = 6.0; \, \gamma = 2.0$ & $0.985 \pm 0.031$ & $0.997 \pm 0.032$ & $0.001 \pm 0.001$ & $0.000 \pm 0.000$ & $0.011 \pm 0.003$ & $0.003 \pm 0.002$ & $0.003 \pm 0.002$ & $0.000 \pm 0.000$ \\
  $\tilde{M}_\bullet = 6.3; \, \gamma = 1.3$ & $0.989 \pm 0.031$ & $0.994 \pm 0.032$ & $0.002 \pm 0.001$ & $0.001 \pm 0.001$ & $0.008 \pm 0.003$ & $0.005 \pm 0.002$ & $0.001 \pm 0.001$ & $0.000 \pm 0.000$ \\
  $\tilde{M}_\bullet = 6.3; \, \gamma = 2.0$ & $0.988 \pm 0.031$ & $0.995 \pm 0.032$ & $0.002 \pm 0.001$ & $0.002 \pm 0.001$ & $0.007 \pm 0.003$ & $0.001 \pm 0.001$ & $0.003 \pm 0.002$ & $0.002 \pm 0.001$ \\
  $\tilde{M}_\bullet = 6.7; \, \gamma = 1.3$ & $0.994 \pm 0.032$ & $0.997 \pm 0.032$ & $0.001 \pm 0.001$ & $0.000 \pm 0.000$ & $0.005 \pm 0.002$ & $0.003 \pm 0.002$ & $0.000 \pm 0.000$ & $0.000 \pm 0.000$ \\
  $\tilde{M}_\bullet = 6.7; \, \gamma = 2.0$ & $0.993 \pm 0.032$ & $0.995 \pm 0.032$ & $0.002 \pm 0.001$ & $0.002 \pm 0.001$ & $0.003 \pm 0.002$ & $0.002 \pm 0.001$ & $0.002 \pm 0.001$ & $0.001 \pm 0.001$ \\
  $\tilde{M}_\bullet = 7.0; \, \gamma = 1.3$ & $0.995 \pm 0.032$ & $0.996 \pm 0.032$ & $0.001 \pm 0.001$ & $0.000 \pm 0.000$ & $0.004 \pm 0.002$ & $0.004 \pm 0.002$ & $0.000 \pm 0.000$ & $0.000 \pm 0.000$ \\
  $\tilde{M}_\bullet = 7.0; \, \gamma = 2.0$ & $0.993 \pm 0.032$ & $0.997 \pm 0.032$ & $0.001 \pm 0.001$ & $0.001 \pm 0.001$ & $0.003 \pm 0.002$ & $0.002 \pm 0.001$ & $0.002 \pm 0.001$ & $0.000 \pm 0.000$ \\
  \midrule
  $N_\mathrm{MC}=10^4$ \\
  \midrule
  $\tilde{M}_\bullet = 4.0; \, \gamma = 2.0$ & $0.683 \pm 0.008$ & $0.966 \pm 0.010$ & $0.032 \pm 0.002$ & $0.010 \pm 0.001$ & $0.227 \pm 0.005$ & $0.021 \pm 0.001$ & $0.059 \pm 0.002$ & $0.003 \pm 0.001$ \\
  $\tilde{M}_\bullet = 6.6; \, \gamma = 1.3$ & $0.989 \pm 0.010$ & $0.994 \pm 0.010$ & $0.002 \pm 0.000$ & $0.001 \pm 0.000$ & $0.007 \pm 0.001$ & $0.004 \pm 0.001$ & $0.001 \pm 0.000$ & $0.001 \pm 0.000$ \\
  \bottomrule
\end{tabular}
\caption{ Fractions of outcomes from the Monte Carlo calculations. The fractions are based on $N_\mathrm{MC}$ simulations for each parameter combination, with and without the inclusion of VRR\@. Each row corresponds to a different combination of $\tilde{M}_\bullet \equiv \log_{10}(M_\bullet/\msun)$ and $\gamma$, indicated in the first column. We show results for the small set of simulations ($N_\mathrm{MC}=10^3$; top part of the table), and the large set ($N_\mathrm{MC}=10^4$; bottom part of the table). Fractions are rounded to three digits, and Poisson errors are given for each entry. }
\label{table:fractions}
\end{table*}

We distinguish between the following channels in the Monte Carlo
simulations.
\begin{enumerate}
\item The binary does not merge, and the binary+MBH system remains outside of
  the semisecular regime at all times (`no interaction').
\item The binary does not merge, but the binary+MBH system does enter the
  semisecular regime at any time during the evolution (`no interaction;
  semisecular').
\item Due to high eccentricities in the inner orbit, stopping condition (1)
  (see \S~\ref{sect:pop_syn:setup:sc}) is met without entering the semisecular
  regime. The binary will merge due to orbital energy loss due to GW emission
  (`merger').
\item Similar to case (3), but here, during any point of the evolution the
  semisecular regime was also entered (`merger; semisecular').
\end{enumerate}
In Table~\ref{table:fractions}, we show the fractions of these channels in the
simulations for the `small' set of simulations ($N_\mathrm{MC}=10^3$; top part
of the table), and the `large' set of simulations ($N_\mathrm{MC}=10^4$; bottom
part of the table). The fractions are given for simulations with VRR included
(`VRR'), and excluded (`No VRR'). Each row corresponds to a different
combination of $\tilde{M}_\bullet \equiv \log_{10}(M_\bullet/\msun)$ and
$\gamma$, indicated in the first column. Poisson errors are given for each
entry. As to be expected, the merger fractions for $M_\bullet=10^4\,\msun$ and
$\gamma=2$ are consistent within the error bars between the `small' and `large'
sets.

A dynamical instability of the binary+MBH system does not occur in the
simulations; therefore, the corresponding fractions are not given in
Table~\ref{table:fractions}. This can be attributed to the constancy of
$e_\oout$ in the simulations (note that the other quantities occurring in
the stability criterion of \citet{2001MNRAS.321..398M} are also constant in our simulations).

We note the following trends in the fractions shown in
Table~\ref{table:fractions}.
\begin{itemize}
\item The merger fractions with VRR included are the largest for
  $M_\bullet=10^4\,\msun$ and $\gamma=2$, in which case
  $f_\mathrm{merge}\approx0.2$. For smaller $M_\bullet$ and $\gamma$, the
  merger fractions are smaller. For $M_\bullet=10^7\,\msun$, the fractions are
  close to zero.
\item Typically, the merger fractions are enhanced by VRR by a factor of up to
  $\sim 10$.
\item Depending on $M_\bullet$ and $\gamma$, for $\sim10\%$ to $\sim 50\%$ of
  the mergers the semisecular regime is entered (i.e., before reaching stopping
  condition 1 in \S~\ref{sect:pop_syn:setup:sc}).
\item Of the non-interacting systems, up to a few per cent enter the
  semisecular regime. The fraction of these systems is a few times larger if
  VRR is included.
\end{itemize}

\subsubsection{Orbital distributions}
\label{sect:pop_syn:results:orbits}
In this section, we focus on the properties of the systems that undergo the
outcomes described in \S~\ref{sect:pop_syn:results:fractions}, specializing to
the `large' set of simulations ($N_\mathrm{MC}=10^4$) with
$M_\bullet=10^4\,\msun$ and $\gamma=2$. The results are qualitatively similar
for the large GC-like set ($M_\bullet=4\times 10^6\,\msun$ and $\gamma=1.3$);
however, the statistics in the former simulations are much better because of the
larger merger fractions (see Table~\ref{table:fractions}). Therefore, we focus
on the case $M_\bullet=10^4\,\msun$ and $\gamma=2$.

\begin{figure}
\center
\includegraphics[width=0.52\textwidth]{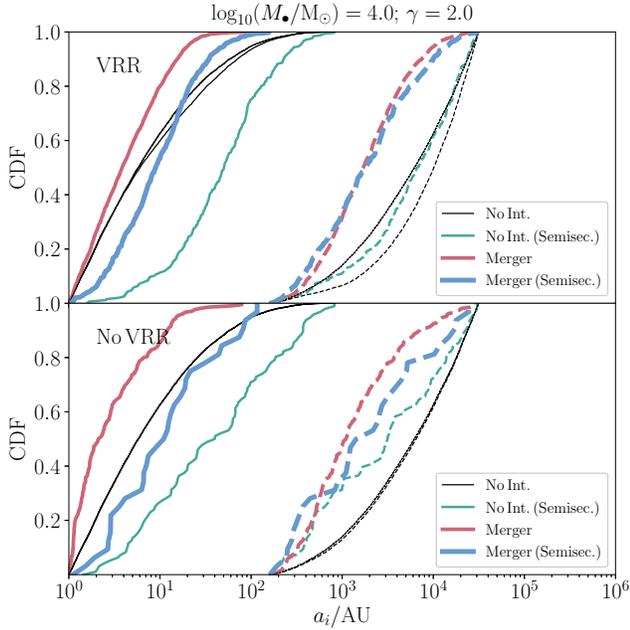}
\caption{\small Cumulative distributions of the inner (solid lines) and outer
  (dashed lines) orbit semimajor axes for the `large' simulations
  ($N_\mathrm{MC}=10^4$), for which $M_\bullet=10^4\,\msun$, and
  $\gamma=2$. The top (bottom) panels apply to the simulations with (without)
  VRR\@. The cases shown are no interaction (black lines), no interaction but
  entering the semisecular regime at any point in the evolution (green lines),
  merger (red lines), and merger while also entering the semisecular regime
  (blue lines). The thin black solid (dashed) lines show the initial
  distributions of the inner (outer) semimajor axes for {\it all} systems
  (i.e., not making a distinction to the outcome).  }
\label{fig:sma_distributions}
\end{figure}

\paragraph{Semimajor axes} In \F~\ref{fig:sma_distributions}, we show the
distributions of the inner (solid lines) and outer (dashed lines) orbit
semimajor axes. In this and in following figures, the top (bottom) panels apply
to the simulations with (without) VRR\@. The cases shown are no interaction
(black lines), no interaction but entering the semisecular regime at any point
in the evolution (green lines), merger (red lines), and merger while also
entering the semisecular regime (blue lines). The thin black solid (dashed)
lines show the initial distributions of the inner (outer) semimajor axes for
{\it all} systems (i.e., not making a distinction with respect to the outcome).

The typical initial inner orbit semimajor axis for the merging systems is
$\sim 10\,\au$. There are few mergers with small $a_\oin$, which can be
attributed to relativistic precession, as noted before in
\S~\ref{sect:time_scales:comp}. Merger systems that also enter the semisecular
regime typically have larger $a_\oin$, which can be understood from
equation~(\ref{eq:semisec}): more compact systems are more susceptible to
entering the semisecular regime. A similar property applies to the
non-interacting semisecular systems. Overall, the semimajor axis distributions
do not vary greatly between the `VRR' and `No VRR' cases.
 
\begin{figure}
   \center
   \includegraphics[width=0.52\textwidth]{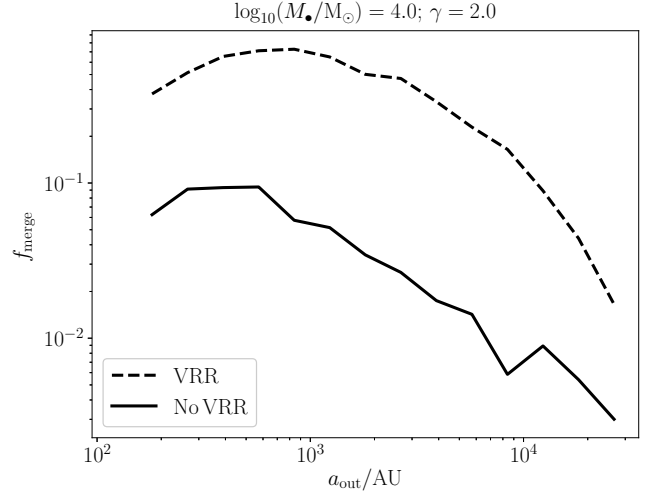}
   \caption{\small Merger fractions (including both the semisecular and non-semisecular mergers) as a function of the outer orbit semimajor axis, $a_\oout$, for the VRR (solid line) and no VRR (dashed line) cases. } 
\label{fig:f_merge_sma}
\end{figure}

Complementary to \F\,\ref{fig:sma_distributions}, we show in \F\,\ref{fig:f_merge_sma} the merger fractions (including both the semisecular and non-semisecular mergers) as a function of the outer orbit semimajor axis, $a_\oout$, for the VRR (solid line) and no VRR (dashed line) cases. With VRR included, the merger fraction peaks around $10^3\,\au$, which corresponds to $\sim 0.1\,r_\mathrm{h}$ for the simulation with $M_\bullet=10^4\,\msun$ and $\gamma=2$. This is consistent with \F\,\ref{fig:timescales}, in the sense that the latter figure shows that for comparable parameters, the regions of interest lie around $r=0.1\,r_\mathrm{h}$. The enhancement relative to the case without VRR is $\sim 10$, and is not strongly dependent on $a_\oout$. 
 
 \begin{figure}
   \center
   \includegraphics[width=0.52\textwidth]{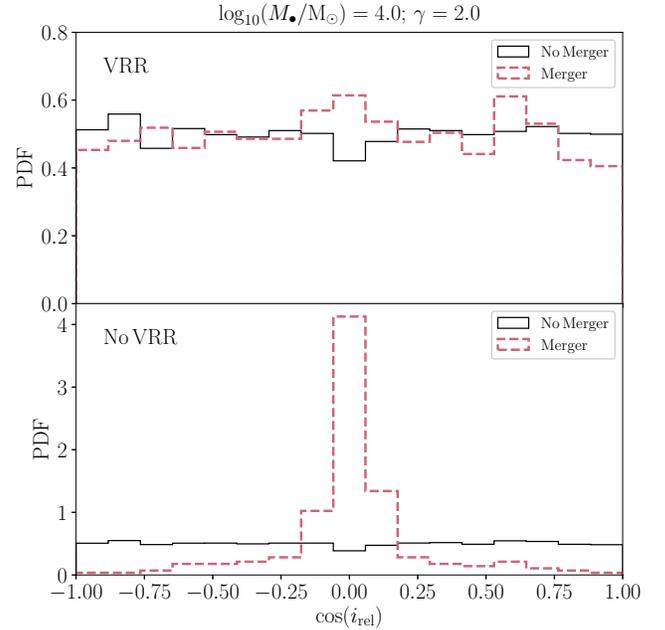}
   \caption{\small Distributions from the Monte Carlo simulations of
     the initial mutual inclinations between the inner and outer orbits,
     arranged by merger outcomes (red dashed lines), and non-merger outcomes
     (black solid lines). Top (bottom) panels apply to simulations with
     (without) VRR\@. The initial distribution of $\cos(i_\mathrm{rel})$ was
     assumed to be flat between -1 and 1. }
\label{fig:incl_distributions}
\end{figure}

\begin{figure}
  \center
  \includegraphics[width=0.52\textwidth]{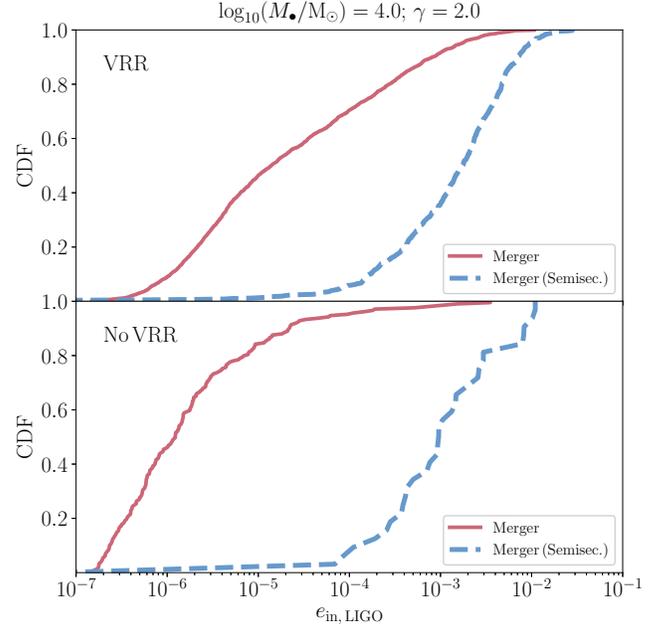}
  \caption{\small Cumulative distributions of the eccentricity of the merging inner
    binaries when they enter the LIGO band, computed using
    equation~(\ref{eq:e_Ligo}). Blue (red) lines show systems that do (not)
    enter the semisecular regime. }
\label{fig:e_distributions}
\end{figure}

\paragraph{Inclinations} The distributions of the initial mutual inclinations
between the inner and outer orbits are shown in
\F~\ref{fig:incl_distributions}. Here, we show distributions for the merging
systems (red dashed lines), and the non-merging systems (black solid lines),
combining results from the systems that enter and that do not enter the
semisecular regime. Initially, random orbital orientations were assumed
(uniform distribution in $\cos i_\mathrm{rel,\,0}$). Without VRR, the inclination
distributions of the semisecular and merger systems are highly peaked around
$90^\circ$. This can be easily understood by noting that these cases are
associated with very high eccentricities (see also
\F~\ref{fig:e_distributions}); therefore, the initial inclination should be
close to $90^\circ$ (see equation~\ref{eq:e_max_LK}). With VRR included, the
distributions for the same channels are not peaked, and broadly distributed
along all inclinations. This can be understood by noting that VRR can induce
high mutual inclinations, which can subsequently lead to high inner orbit
eccentricities through LK evolution. Therefore, no high initial mutual
inclination is required, and this is the key ingredient to the enhanced merger
rates when VRR is taken into account.

\begin{figure}
  \center
  \includegraphics[width=0.52\textwidth]{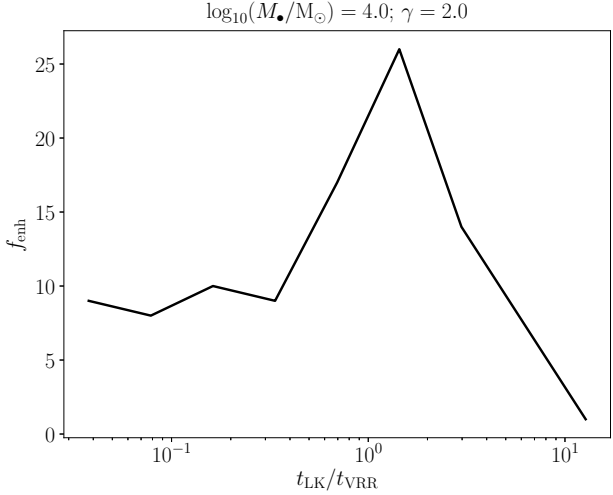}
  \caption{\small The enhancement factor due to VRR of the mergers in the Monte Carlo simulations, i.e., the number of mergers in the case of VRR divided by the number of mergers without VRR, binned in terms of the adiabatic parameter $\adpar\equiv t_\mathrm{LK}/t_\mathrm{VRR}$ (see equation~\ref{eq:adpar}). }
\label{fig:R_distributions}
\end{figure}

\paragraph{Adiabatic parameter}
In \F\,\ref{fig:R_distributions}, we show the enhancement factor due to VRR of the mergers in the Monte Carlo simulations, i.e., the number of mergers in the case of VRR divided by the number of mergers without VRR, binned in terms of the adiabatic parameter $\adpar\equiv t_\mathrm{LK}/t_\mathrm{VRR}$ (see equation~\ref{eq:adpar}). We only include bins in $\adpar$ with more than two mergers in the case without VRR. The enhancement factor is peaked near $\adpar=1$, which is in agreement with the scale-free results of \S\,\ref{sect:dyn}. Some enhancement is already present at $\adpar \sim 0.1$, which is consistent with \F\,\ref{fig:emax_cdfs} in the sense that in the latter figure, the cumulative distribution of the maximum eccentricity is already significantly enhanced from the case without VRR ($\adpar=0$) for $\adpar\sim0.1$. 

\paragraph{Eccentricity at the LIGO band} In \F~\ref{fig:e_distributions}, we
show the distributions of the eccentricity of the merging binaries when they
enter the LIGO band, $e_{\oin,\,\mathrm{LIGO}}$. Here, we compute
$e_{\oin,\,\mathrm{LIGO}}$ by assuming that $e_{\oin,\,\mathrm{LIGO}}$ is
small, such that (using equations~\ref{eq:f_GW} and~\ref{eq:ae_peters})
\begin{align}
\label{eq:e_Ligo}
  e_{\oin,\,\mathrm{LIGO}} \approx {\left [ \frac{1}{\pi f_\mathrm{GW,\,LIGO}} \sqrt{ \frac{G (M_1+M_2)}{c_0^3}  } \right ]}^{19/18},
\end{align}
where $f_\mathrm{GW,\,LIGO} = 20\,\mathrm{Hz}$, and $c_0$ is determined using
equation~\eqref{eq:ae_peters} and the simulation data at the moment of the
merger stopping condition (as shown above, at this point the inner binary is
decoupled). Equation~\eqref{eq:e_Ligo} applies only if
$e_{\oin,\,\mathrm{LIGO}}$ is small; we find a posteriori that this is indeed
the case --- the true GW frequency at $e_{\oin,\,\mathrm{LIGO}}$ in our
simulations deviates from $f_\mathrm{GW,\,LIGO}$ by no more than a few per
cent.

The typical eccentricity when entering the LIGO band is small. The median
value in the VRR case is $\sim 10^{-5}$, with maximum values of $\sim 10^{-2}$
for the systems that also entered the semisecular regime (blue lines in
\F~\ref{fig:e_distributions}). The latter case is associated with very high
eccentricities, explaining the typically larger values of
$e_{\oin,\,\mathrm{LIGO}}$. The values of $e_{\oin,\,\mathrm{LIGO}}$ are
typically larger if VRR is included. This can be explained by noting that the
typical excited eccentricities in the VRR case are larger compared to the
`pure' LK case.

\begin{figure}
  \center
  \includegraphics[width=0.52\textwidth]{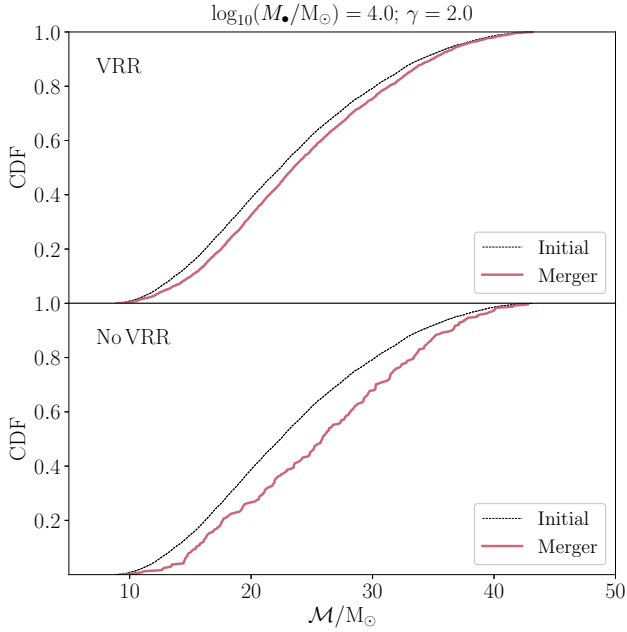}
  \caption{\small Cumulative distributions of the Chirp masses,
    $\mathcal{M} = {(M_1 M_2)}^{3/5}/{(M_1+M_2)}^{1/5}$, for the mergers in the
    Monte Carlo simulations (solid red lines). The thin black dotted lines
    show the initial distributions. Top (bottom): with (without) VRR\@. }
\label{fig:mass_distributions}
\end{figure}

\begin{figure}
  \center
  \includegraphics[width=0.52\textwidth]{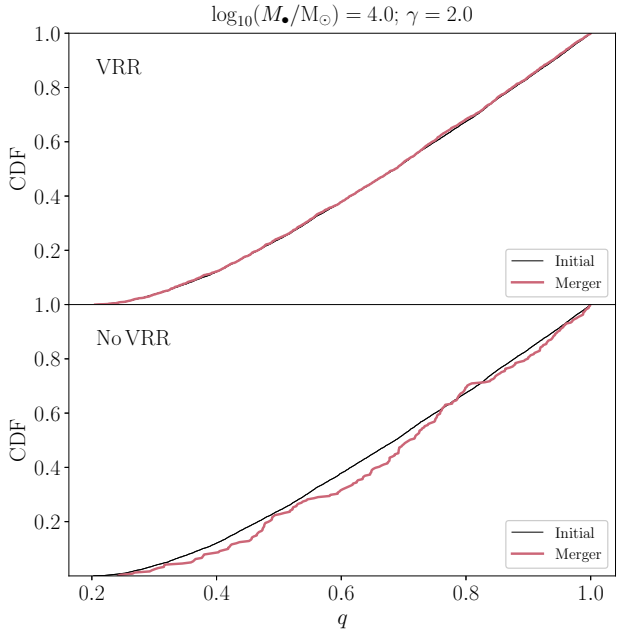}
  \caption{\small Cumulative distributions of the binary mass ratio, 
    $q\equiv M_2/M_1$, for the mergers in the
    Monte Carlo simulations (solid red lines). The thin black dotted lines
    show the initial distributions. Top (bottom): with (without) VRR\@. }
\label{fig:mass_ratio_distributions}
\end{figure}

\paragraph{Masses} In \F~\ref{fig:mass_distributions}, we show the
distributions of the Chirp mass,
$\mathcal{M} = {(M_1 M_2)}^{3/5}/{(M_1+M_2)}^{1/5}$, for the merging systems in
the simulations. The initial distributions for all systems are shown with the
thin black solid lines. The merger systems have slightly higher Chirp masses
compared to all systems, although the differences are small --- it is unlikely
that these differences have observational implications.

We show a similar figure for the mass ratios $q\equiv M_2/M_1$ 
in \F\,\ref{fig:mass_ratio_distributions}. The mergers show no discernible dependence on $q$.

\begin{figure}
  \center
  \includegraphics[width=0.52\textwidth]{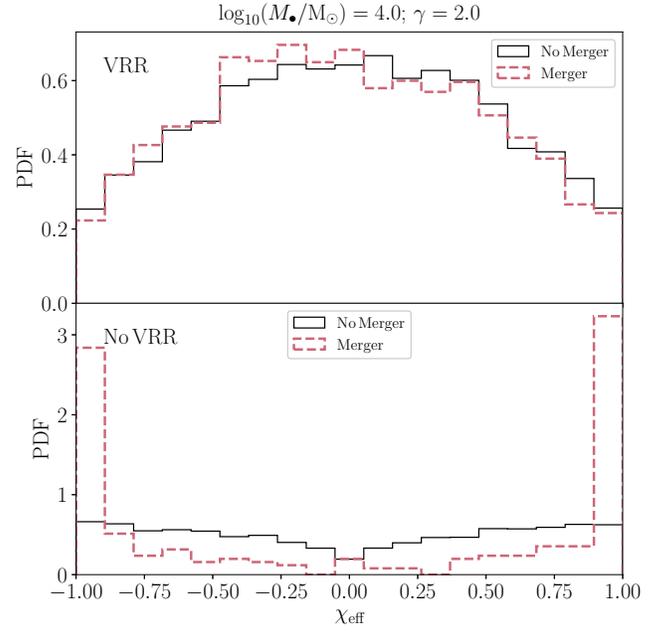}
  \caption{\small Distributions of the final $\chi_\mathrm{eff}$ (see
    equation~\ref{eq:chi_eff}) for the four outcomes in the Monte Carlo simulations.}
\label{fig:chi_eff_distributions}
\end{figure}

\paragraph{Spins} Finally, we show in \F~\ref{fig:chi_eff_distributions} the
distributions of the projected spin parameter, i.e.,
\begin{align}
\label{eq:chi_eff}
\chi_\mathrm{eff} = \frac{M_1 \chi_1 \cos(\theta_1) + M_2 \chi_2 \cos(\theta_2)}{M_1+M_2},
\end{align}
where $\chi_i$ is the normalized spin parameter of star $i$, and $\theta_i$ is
the angle between the spin vector of star $i$ and the inner orbit
angular-momentum vector. For simplicity, we assume that the two compact objects
are maximally spinning, i.e., $\chi_1=\chi_2=1$. The distributions of
$\chi_\mathrm{eff} $ are shown at the end of the simulations for merging and
non-merging systems (both including the semisecular systems). Note that for the
merger systems, this corresponds to a point in the evolution in which the inner
binary is decoupled from the MBH and the stellar cluster. Near the LIGO band,
 the higher-order spin-spin and spin-orbit terms will become important and change
  the individual spin orbit angles. However, $\chi_\mathrm{eff}$ does not change 
  during this process at the 2.5PN level. The initial spin-orbit angles were assumed to be zero.
  We note that the spin dynamics of compact objects in triple systems (without taking into account
the effects of VRR) have been considered by a number of authors
\citep{2017ApJ...846L..11L,2017arXiv171107142A,2018arXiv180503202L}.

Without VRR, the distribution of $\chi_\mathrm{eff}$ for the merging systems is
peaked near $-1$ and $+1$. With VRR included, the distributions of
$\chi_\mathrm{eff}$ are broadly distributed between $-1$ and $+1$, with a peak around
0, indicative of random spin-orbit alignment. This can be
ascribed to the coupling of LK oscillations with VRR: VRR continuously
adjusts the outer orbital angular momentum vector independent of the initial
mutual inclination $i_\mathrm{rel}$, and these changes are transmitted to the
inner orbit through the LK mechanism, producing spin-orbit misalignment for both 
merging and non-merging systems. Given that the initial spin-orbit alignment was
assumed to be zero, the top panel of \F\,\ref{fig:chi_eff_distributions}
shows that LK-VRR is very effective at producing random spin-orbit orientations.

\subsubsection{Stopping times}
\label{sect:pop_syn:results:stop_times}

\begin{figure}
\center
\includegraphics[width=0.52\textwidth]{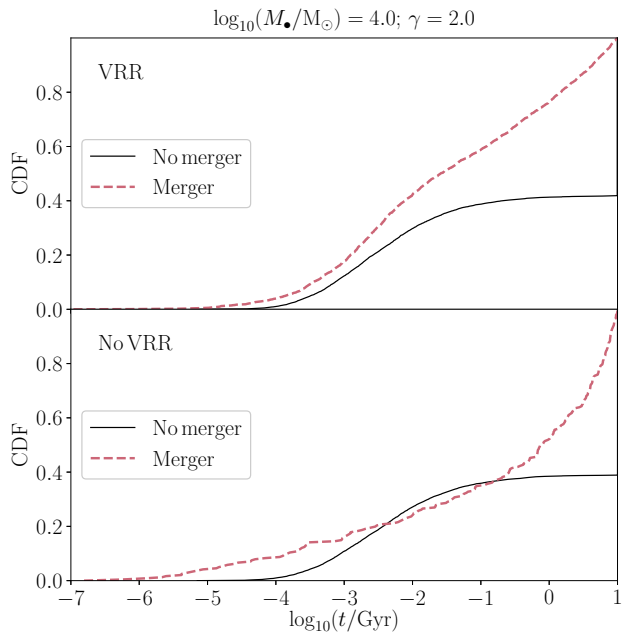}
\caption{\small Cumulative distributions of the stopping times in the Monte Carlo
  simulations arranged by the outcomes (merger: red dashed lines; no merger:
  solid black lines).}
\label{fig:end_time_distributions}
\end{figure}

In \F~\ref{fig:end_time_distributions}, we show the distributions of the
stopping times arranged by the outcomes. For the non-merging systems, the
stopping times are either set by the evaporation time-scale or 10 Gyr; for the
mergers, the stopping time is the time when GW emission started to dominate
over the LK dynamics; the true merging time is very close to this (see
\S~\ref{sect:pop_syn:setup:sc}). The merger times in the simulations span a
wide range between $\sim0.1\,\mathrm{Myr}$, and $10\,\mathrm{Gyr}$. With VRR
included, the median stopping time is $\sim 100\, \mathrm{Myr}$. The late
mergers are associated with a slower type of evolution in which the semimajor
axis decreases only slightly at high eccentricities, until finally the 2.5PN
terms start to dominate and the inner binary becomes decoupled.

\begin{figure}
  \center
  \includegraphics[width=0.52\textwidth]{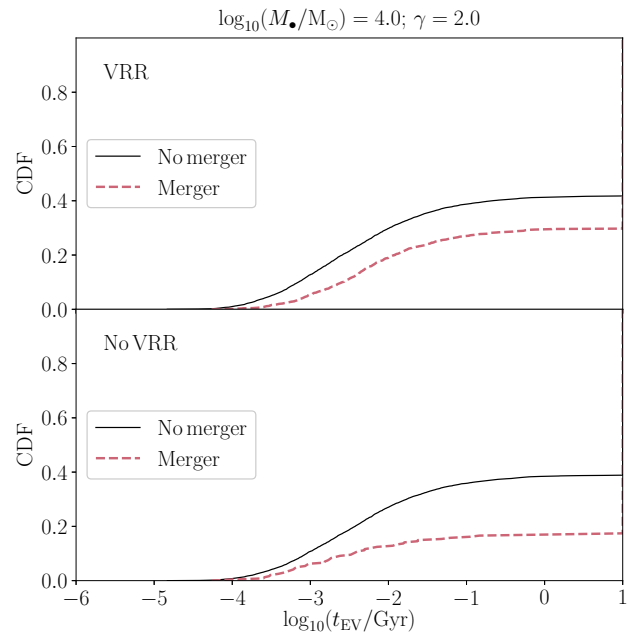}
  \caption{\small Cumulative distributions of the evaporation times (see
    equation~\ref{eq:t_EV}) in the Monte Carlo simulations arranged by
    the outcomes (merger: red dashed lines; no merger: solid black lines).}
\label{fig:evaporation_time_distributions}
\end{figure}

Some, but not all mergers are limited by binary evaporation. This is
illustrated in \F~\ref{fig:evaporation_time_distributions}, in which the
distributions are shown of the evaporation time-scale arranged by the
outcomes. About 60\% of the mergers with VRR included have evaporation
time-scales longer than 10 Gyr.

\subsubsection{Delay-time distributions}
\label{sect:pop_syn:results:DTD}

\begin{figure}
  \center
  \includegraphics[width=0.52\textwidth]{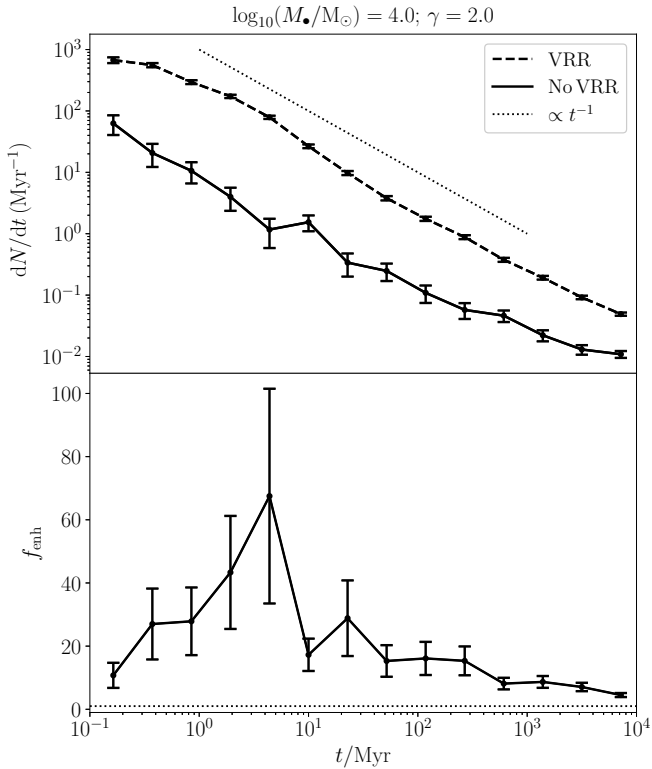}
  \caption{\small Top panel: DTDs of the mergers in the `large' Monte Carlo simulation set with $M_\bullet=10^4\,\msun$ and
    $\gamma=2$. Dashed (solid) lines apply to the simulations with (without)
    VRR\@. Error bars are based on Poisson statistics. The black dotted line
    shows a $\propto t^{-1}$-dependence. Bottom panel: the enhancement factor,
    i.e., the ratio of the DTDs for the cases with and without VRR, as a
    function of time. The horizontal black dashed line shows an enhancement
    factor of 1.}
\label{fig:DTD}
\end{figure}
From the simulations, we determine the delay-time distributions (DTDs), i.e.,
the number of events per unit time. We consider the mergers in the `large' set
of simulations with $M_\bullet=10^4\,\msun$ and $\gamma=2$, and include the
mergers in which the semisecular regime was entered. The resulting DTDs are
shown in the top panel of \F~\ref{fig:DTD}. Dashed (solid) lines apply to the
simulations with (without) VRR\@. For both cases with and without VRR, the DTDs
are approximately inversely proportional with time. However, the absolute
normalization of the DTDs is significantly higher with VRR included; the bottom
panel in \F~\ref{fig:DTD} shows the enhancement factor of the DTDs of the cases
with VRR relative to no VRR\@. The enhancement factor is a function of time; it
peaks at $\sim 60$ around 4 Myr, and drops toward $\approx 5$ at
$10^4\,\mathrm{Myr}$.

\section{Discussion}
\label{sect:discussion}

\subsection{Uncertainties in the integrations}
\label{sect:discussion:uncertainties}
As mentioned in \S~\ref{sect:time_scales}, galactic nuclei are complex
dynamical environments. Our Monte Carlo simulations
(\S~\ref{sect:pop_syn}), although the first to include the effects of VRR in
the context of the secular (orbit-averaged) LK problem, did not include all
effects. Here, we comment on some of the processes that we did not consider,
and should be included in future work.

As mentioned in \S~\ref{sect:introduction}, the torques exerted by the stellar cluster not only
affect the orientation of an orbit around an MBH, but also the magnitude of its
angular momentum. This process, known as SRR, was not included in the
simulations. We argued in \S~\ref{sect:time_scales} that the SRR time-scale is
typically longer than the VRR time-scale. However, the merging of the inner
binary can occur after many LK cycles, corresponding typically to times
$\gg t_\mathrm{VRR}$ (since typically $\adpar \sim1$ to reach high
eccentricities, which are necessary for merging systems), and SRR can affect
the outer orbit eccentricity on these longer time-scales. This could be
important, since $e_\mathrm{out}$ changes both the efficiency of VRR and the LK
process. More dramatically, an increased $e_\mathrm{out}$ could bring the
binary sufficiently close to the MBH for the former to be tidally disrupted
\citep{1988Natur.331..687H}.

Other effects that were neglected in the simulations are changes of the outer
orbital parameters due to NRR and due to dynamical friction (DF). NRR in particular could be important if 
the mass function of the stellar background is top heavy, significantly reducing the
NRR time-scale (see equation~\ref{eq:NRR}); this is an effect that we have 
ignored. DF could be important if the binary is much more massive than the typical
background star, which is the case in our simulations (on average, $M_\mathrm{bin}\equiv
M_1+M_2=60\,\msun$, such that $M_\mathrm{bin}/m_\star = 60$). The DF time-scale can be estimated
as $t_\mathrm{DF} \sim t_\mathrm{NR} \, (m_\star/M_\mathrm{bin}) \sim t_\mathrm{NR}/60$ \citep{1987degc.book.....S,2009ApJ...697.1861A},
implying that DF could be important in our simulations. 

NRR and DF can affect the binary in several ways. First, a change in the outer binary semimajor axis and
eccentricity can bring the binary into a different dynamical regime, since
$\adpar$ (see equation~\ref{eq:adpar}) is a function of both $a_\oout$ and
$e_\oout$. Also, the binary can migrate to regions with shorter evaporation
time-scales, such that the binary may not survive until a merger is
triggered. Alternatively, the binary can be brought too close to the MBH and be
tidally disrupted.

\subsection{Estimates of absolute merger rates}
\label{sect:discussion:abs}
Although the focus of this paper is on the {\it relative enhancement} of VRR on the
rates of LK-induced mergers of binaries in galactic nuclei,
 we here estimate the absolute BH-BH binary merger rates based on our
Monte Carlo simulations. A large number of 
assumptions must be made in order to make these estimates, and we emphasize
that the numbers given here are extremely uncertain. The uncertainties in the absolute
rates can be considered to be at least one order of magnitude, and up to several orders of magnitude. 

\subsubsection{GC}
\label{sect:discussion:abs:GC}
First, we consider our fiducial GC model ($M_\bullet = 4 \times 10^6\,\msun$; $\gamma=1.3$). 
We follow similar assumptions as \citet{2017ApJ...846..146P}. Let
$n_\mathrm{gal}$ be the number density of Milky-Way-like galaxies,
$\Gamma_\mathrm{CO}$ be the formation rate of compact objects (number per unit
time), $f_\mathrm{CO}$ the fraction of stars that evolve to compact object
binaries (this includes the binary fraction), $f_\mathrm{merge}$ the merger fractions in the Monte Carlo
 simulations (\S~\ref{sect:pop_syn}), and $f_\mathrm{calc}$ the fraction of calculated
systems relative to the underlying binary population. The merger rate (number per unit
time per unit volume) is then estimated by
\begin{align}
\label{eq:rates}
  \Gamma \sim n_\mathrm{gal} \Gamma_\mathrm{CO} f_\mathrm{CO} f_\mathrm{merge} f_\mathrm{calc}.
\end{align}
We adopt a number density of galaxies of
$n_\mathrm{gal} \approx 0.02 \, \mathrm{Mpc^{-3}}$
\citep{2005ApJ...620..564C,2008ApJ...675.1459K}.

We use simple arguments to estimate the compact object formation rate (for more detailed studies,
see, e.g., \citealt{2015ApJ...799..185A,2016ApJ...823..137A}). 
Assuming constant star formation, \citet{2010MNRAS.402..519L} find that
$~\sim 10^4$ ($~\sim 10^5$) BHs are formed for $10^6\,\msun$ of mass in
stars and stellar remnants, assuming a canonical (top-heavy) initial-mass
function (IMF). Scaling these numbers to the GC, for which the total mass of
stars formed in $\sim 10\,\mathrm{Gyr}$ is $\sim 10^7\,\msun$
\citep{2014AA...566A..47S,2017MNRAS.466.4040F,2018AA...609A..27S}, this implies
that the rate of compact object formation in the GC is
$\Gamma_\mathrm{CO} \sim 10^{-5}\,\mathrm{yr^{-1}}$
($\Gamma_\mathrm{CO} \sim 10^{-4}\,\mathrm{yr^{-1}}$) assuming a canonical
(top-heavy) IMF\@. 

Based on the \textsc{BSE} binary population synthesis code
\citep{2002MNRAS.329..897H}, \citet{2017ApJ...846..146P} find that the fraction of
stars forming compact object binaries is $0.025$ ($0.045$) for a canonical (top-heavy)
IMF\@. Here, the canonical (top-heavy) IMF is defined as
$\mathrm{d}N/\mathrm{d}m \propto m^{-2.3}$
($\mathrm{d}N/\mathrm{d}m \propto m^{-1.7}$). In these fractions, which are related to $f_\mathrm{CO}$,
\citet{2017ApJ...846..146P} implicitly assumed a MS binary fraction
of unity. The underlying distributions of the binary properties in our simulations are different
from those of \citet{2017ApJ...846..146P}, implying that the appropriate values of $f_\mathrm{CO}$
are not necessarily the same as in \citet{2017ApJ...846..146P}. Nevertheless, we let $f_\mathrm{CO}$
be motivated by the results of \citet{2017ApJ...846..146P}, and so we set $f_\mathrm{CO}=0.25$
 ($f_\mathrm{CO}=0.05$) for the canonical (top-heavy) IMF.

From Table~\ref{table:fractions},
we adopt a merger fraction of $f_\mathrm{merge} = 0.008$ and
$f_\mathrm{merge} = 0.005$ for the cases with and without VRR, respectively (we
combine the merger fractions in the semisecular and non-semisecular regimes). We note
that these merger fractions are based on simulations in which we assumed 
a Salpeter mass function for the stellar cluster
(i.e., $m_\star\equiv \langle m^2 \rangle ^{1/2} = 1\,\msun$, see \S\,\ref{sect:time_scales}). 
With a top-heavy mass function, the merger fractions could be different since $m_\star$ affects
the VRR and evaporation time-scales. Here, we neglect this complication, and assume that
$f_\mathrm{merge}$ is the same for the standard and top-heavy mass functions.

We estimate the calculated fraction, $f_\mathrm{calc}$, by computing the fraction of
orbits in our Monte Carlo simulations relative to the entire population, assuming that
the semimajor axis distribution of the latter ranges between $\sim10^{-2}$ and $\sim10^6\,\au$. For the GC model, the
simulated binaries have semimajor axes ranging between $\sim1$ and $\sim10^3\,\au$. Given that
the distribution in $a_\oin$ is assumed to be flat in $\ln(a_\oin)$, this implies that 
$f_\mathrm{calc} \sim \ln(10^3)/\ln(10^6) = 1/2$. 

With the above assumptions, we find merger rates of
$\Gamma_\mathrm{VRR} = 0.02\,\mathrm{Gpc^{-1}\,yr^{-1}}$
($\Gamma_\mathrm{VRR} = 0.4\,\mathrm{Gpc^{-1}\,yr^{-1}}$) assuming a canonical
(top-heavy) IMF if VRR is included. Without VRR, we find
$\Gamma_\mathrm{No\,VRR} \simeq 0.013\,\mathrm{Gpc^{-1}\,yr^{-1}}$
($\Gamma_\mathrm{No\,VRR} = 0.25\,\mathrm{Gpc^{-1}\,yr^{-1}}$) assuming a
canonical (top-heavy) IMF\@.

For reference, predictions of the BH merger rate for globular clusters include
2-20 $\mathrm{Gpc^{-3}\,yr^{-1}}$ \citep{2016PhRvD..93h4029R}, 
$>6.5 \,\mathrm{Gpc^{-3}\,yr^{-1}}$ \citep{2017MNRAS.469.4665P},
$>5.4 \,\mathrm{Gpc^{-3}\,yr^{-1}}$ \citep{2017MNRAS.464L..36A}, and 
15-100 $\mathrm{Gpc^{-3}\,yr^{-1}}$ \citep{2018arXiv180602351F}.
For nuclear star clusters without MBHs, the rates have been predicted to be
$\sim 1 \, \mathrm{Gpc^{-3}\,yr^{-1}}$ \citep{2016ApJ...831..187A}. The BH-BH
merger rate inferred by LIGO is 12--213 $\mathrm{Gpc^{-1}\,yr^{-1}}$
\citep{2017PhRvL.118v1101A}.

Our numbers can be considered to be low compared to the above numbers. This indicates
that VRR cannot sufficiently increase the merger rates of BH-BH binaries enough
to make a significant contribution to the LIGO rate. However, the enhancement
 of $f_\mathrm{merge}$ due to VRR is small for GC-like galactic nuclei (i.e., a factor of
$0.008/0.005 = 1.6$). We have shown that the
enhancement of $f_\mathrm{merge}$ is much more significant for galactic nuclei
with lower MBH masses (up to $\sim10$ for $M_\bullet = 10^4\,\msun$ and
$\gamma=2$).

\subsubsection{Other galactic nuclei}
\label{sect:discussion:abs:other}
Here, we make crude estimates for the merger rates in other galactic nuclei
based on our `small' set of simulations ($N_\mathrm{MC}=10^3$). The largest
merger fractions in the simulations occur for low MBH masses of $10^4\,\msun$;
the existence of galactic nuclei with MBHs with such low masses is highly
uncertain. Here, we assume that galactic nuclei with MBHs of such low mass
exist, and estimate their number density as follows. 

The luminosity function
(number density per unit luminosity) for galaxies with MBH masses
$\lesssim 10^8\, \msun$ is approximately $\phi(L) \propto L^{-1}$
\citep{2002AJ....124.3035A}. Since the Faber-Jackson relation, approximately
$L\propto \sigma_\star^4$ \citep{1976ApJ...204..668F}, and the
$M_\bullet-\sigma_\star$ relation, approximately
$M_\bullet \propto \sigma_\star^4$ \citep{2001ApJ...547..140M}, have
approximately the same scaling with $\sigma_\star$, we assume that
$L\propto M_\bullet$. This implies that the number density of MBHs per unit
mass scales as $\phi(M_\bullet) \propto M_\bullet^{-1}$
(\citealt{2002AJ....124.3035A}; we also assume an MBH occupation fraction of
unity). This implies that the $M_\bullet$-integrated number density is
\begin{align}
n_\mathrm{gal} \propto \int \phi(M_\bullet) \, \mathrm{d} M_\bullet \propto \log(M_\bullet),
\end{align}
i.e., the dependence of $n_\mathrm{gal}$ on $M_\bullet$ is weak
(logarithmic). Therefore, for simplicity, we assume that $n_\mathrm{gal}$ is
completely independent of $M_\bullet$, and set
$n_\mathrm{gal} = 0.02\, \mathrm{Mpc^{-3}}$
\citep{2005ApJ...620..564C,2008ApJ...675.1459K}, as for the GC case in
\S~\ref{sect:discussion:abs:GC}.

The compact object formation rate, $\Gamma_\mathrm{CO}$, is also highly
uncertain. We make two limiting assumptions. (1) $\Gamma_\mathrm{CO}$ is
weakly dependent of $M_\bullet$, which would be appropriate if the compact objects
are supplied to the galactic nucleus by NRR. In particular,
assuming the $M_\bullet$-$\sigma$ relation in the form $M_\bullet \propto \sigma^\beta$, 
one can show that (e.g., \citealt{2017NatAs...1E.147A})
\begin{align}
  \Gamma_\mathrm{CO} = \frac{\gamma_\mathrm{CO} N_\star(r_\mathrm{h}) }{t_\mathrm{NRR}(r_\mathrm{h})} \propto M^{\frac{3-\beta}{\beta}}.
\end{align}
The latter is weakly dependent on $M_\bullet$ for $4<\beta<5$, which is typically assumed for the 
$M_\bullet$-$\sigma$ relation. For simplicity, we ignore the dependence of $\Gamma_\mathrm{CO}$ on 
$M_\bullet$. (2)
$\Gamma_\mathrm{CO}$ scales linearly with $M_\bullet$, which is more
appropriate if the supply rate is set by star formation. In case (1), we
calibrate $\Gamma_\mathrm{CO}$ to the GC, and we estimate
\begin{align}
  \Gamma_\mathrm{CO} = \frac{\gamma_\mathrm{CO} N_\star(r_\mathrm{h}) }{t_\mathrm{NRR}(r_\mathrm{h})} \sim \frac{0.1 \times 10^7}{10^{10}\,\mathrm{yr}} = 10^{-4} \, \mathrm{yr^{-1}},
\end{align}
where $\gamma_\mathrm{CO} = 0.1$ is the fractional number of compact objects
formed for a stellar population assuming a top-heavy IMF
\citet{2010MNRAS.402..519L}, $N_\star(r_\mathrm{h}) \sim 10^7$ is the number of
stars at the radius of the sphere of influence
\citep{2014AA...566A..47S,2017MNRAS.466.4040F,2018AA...609A..27S}, and
$t_\mathrm{NRR}(r_\mathrm{h}) \sim 10^{10}\,\mathrm{yr}$ is the NRR time-scale
at the radius of the sphere of influence (e.g.,
\citealt{2010ApJ...718..739M}). In case (2), we normalize
$\Gamma_\mathrm{CO} \propto M_\bullet$ by the GC top-heavy value that we
adopted in \S~\ref{sect:discussion:abs:GC}, i.e.,
\begin{align}
\Gamma_\mathrm{CO} = \left ( \frac{M_\bullet}{4\times 10^6\,\msun} \right ) \,10^{-4} \, \mathrm{yr^{-1}}.
\end{align}
Note that, for the GC ($M_\bullet = 4\times 10^6\,\msun$), the two assumptions give the same value of $\Gamma_\mathrm{CO} $. 

We assume that the compact object binary formation efficiency
$f_\mathrm{CO}$ is independent of $M_\bullet$, and set it to
$f_\mathrm{CO} = 0.05$ for a top-heavy IMF, to be consistent with the GC case (\S\,\ref{sect:discussion:abs:GC}).

Strictly speaking, the calculated fraction $f_\mathrm{calc}$ depends on $M_\bullet$ since the latter sets the scale of
the sphere of influence, thereby the typical outer orbit semimajor axes, and thereby the upper value of
$a_\oin$ in the simulations (due to the requirement of dynamical stability). However, this is not a strong effect, with
the upper value of $a_\oin$ decreasing to $\sim 10^2\,\au$ for $M_\bullet=10^4\,\msun$ (see, e.g.,
 \F\,\ref{fig:sma_distributions}), implying that $f_\mathrm{calc} = 1/3$. For simplicity, we here set 
 $f_\mathrm{calc}$ to $1/2$, independent of $M_\bullet$. 

\begin{figure}
  \center
  \includegraphics[width=0.52\textwidth]{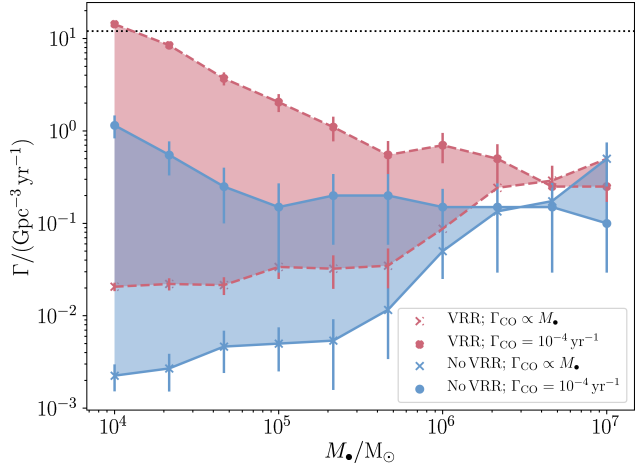}
  \caption{\small Rough estimates of the merger rates based on
    equation~(\ref{eq:rates}) and the Monte Carlo calculations of
    \S~\ref{sect:pop_syn} as a function of $M_\bullet$ for the cases with and
    without VRR (dashed red lines and solid blue lines, respectively). Here, we
    either assume that the compact object formation efficiency
    $\Gamma_\mathrm{CO}$ is independent of $M_\bullet$ and equal to
    $10^{-4} \, \mathrm{yr^{-1}}$ (bullets), or scales linearly with
    $M_\bullet$, and normalized such that
    $\Gamma_\mathrm{CO} = [M_\bullet/(4\times 10^6\,\msun)]\,10^{-4} \,
    \mathrm{yr^{-1}}$ (crosses). The areas in between these cases are shaded
    with red and blue colors for the simulations with VRR included and not
    included, respectively. The number density of galaxies is set to a constant
    $n_\mathrm{gal} = 0.02 \, \mathrm{Mpc^{-3}}$, and the compact object
    formation efficiency $f_\mathrm{CO}$ is assumed to be independent of
    $M_\bullet$, and set it to $f_\mathrm{CO} = 0.045$. The horizontal black
    dotted line shows the lower LIGO limit of
    $12 \, \mathrm{Gpc^{-3}\,yr^{-1}}$ \citep{2017PhRvL.118v1101A}. The error
    bars are based on Poisson statistics from the Monte Carlo
    simulations, i.e., they reflect the uncertainty in $f_\mathrm{merge}$, but
    not the (much larger) errors in the other factors in
    equation~(\ref{eq:rates}), in particular $\Gamma_\mathrm{CO}$. }
\label{fig:rates_summary}
\end{figure}

With these (strong and uncertain) assumptions, and the data from
Table~\ref{table:fractions} for the merger fractions $f_\mathrm{merge}$ (adding
the fractions for the semisecular and non-semisecular mergers), we plot in
\F~\ref{fig:rates_summary} the expected merger rates as a function of
$M_\bullet$, for the cases with and without VRR\@. For
$M_\bullet \leq 4\times 10^6\,\msun$, assumption (1) regarding
$\Gamma_\mathrm{CO}$ gives the highest rates, and in this case we take the data
from the simulations for $\gamma = 2$ (giving an upper limit on the merger
rates); assumption (2) gives much lower rates, and in this case we take data
from the simulations for $\gamma=1.3$ (giving a lower limit). The areas in
between these limits are indicated with shaded regions in
\F~\ref{fig:rates_summary}.

We emphasize that the rates given in \F~\ref{fig:rates_summary} are highly
speculative (note that the error bars reflect the Poisson uncertainty in
$f_\mathrm{merge}$, but not the much larger uncertainties in the other factors in
equation~\ref{eq:rates}, in particular $\Gamma_\mathrm{CO}$). Nevertheless,
taken on face value, \F~\ref{fig:rates_summary} indicates that the merger rates
are low. For galactic nuclei with low MBH masses (less than a few times $10^4\,\msun$), 
the rates are higher, in part due the increased enhancement by VRR. However, 
the existence of such galactic nuclei is not established, and the
most optimistic rates still barely reach the lower limit of the rate inferred by LIGO, 
$12 \, \mathrm{Gpc^{-3}\,yr^{-1}}$, and are an order of magnitude lower than the upper LIGO limit,
$213  \, \mathrm{Gpc^{-3}\,yr^{-1}}$ \citep{2017PhRvL.118v1101A}.

\subsection{Comparison to previous work}
\label{sect:discussion:comp}
Here, we compare the results from our Monte Carlo simulations to
previous work focussing on the dynamics of binaries in galactic nuclei. The
case without VRR has been considered by a number of authors (e.g.,
\citealt{2012ApJ...757...27A,2014ApJ...781...45A,2015ApJ...799..118P,2016MNRAS.460.3494S,
2016ApJ...831..187A,2017ApJ...846..146P,2018ApJ...856..140H}). For
our GC-like model, we find a merger fraction of $f_\mathrm{merge} = 0.005$
(combining the semisecular and non-semisecular outcomes). This is similar
(within a factor of few) to the fraction of \citet{2017ApJ...846..146P}, who
find $f_\mathrm{m} = 0.0017$ (their table 1, with $\epsilon_z = 0.001$).

On the other hand, \citet{2018ApJ...856..140H} find much higher merger
fractions, of up to $f_\mathrm{merge} = 0.15$ for their nominal GC model, which
are higher than the merger fractions of \citet{2017ApJ...846..146P} and our
merger fraction by two orders of magnitude. This large difference can be
ascribed to different initial conditions. About half of the binaries
in \citet{2018ApJ...856..140H} merge due the gravitational wave emission only
 (i.e., even in absence of the
torque of the MBH). Moreover, \citet{2018ApJ...856..140H} considered binaries
much closer to the MBH ($a_\mathrm{out}/r_\mathrm{h} \sim 10^{-4}$ to $\sim 10^{-2}$)
for which the octupole-order terms are important, in contrast to our simulations.

The merger rate predicted by \citet{2018ApJ...856..140H} is $\sim 1$-$3\,\mathrm{Gpc^{-1}\,yr^{-1}}$, 
which is one (two) orders of magnitude larger than our GC rate estimate without VRR assuming a
 top heavy (canonical) IMF. This can be ascribed to the following.
 \begin{enumerate}
 \item The merger fraction adopted by \citet{2018ApJ...856..140H} is $f_\mathrm{merge}=0.15$,
  which is $0.15/0.005=30$ times larger than our adopted fraction. 
 \item The `conversion' factor from merger fraction to rate adopted by
  \citet{2018ApJ...856..140H} is $\Gamma/f_\mathrm{merge} = 2/0.15 \,\mathrm{Gpc^{-1}\,yr^{-1}} \simeq 13\,\mathrm{Gpc^{-1}\,yr^{-1}}$.
   We assumed a `conversion' factor of $\Gamma/f_\mathrm{merge} = 2.5\,\mathrm{Gpc^{-1}\,yr^{-1}}$
    and $\Gamma/f_\mathrm{merge} = 50 \,\mathrm{Gpc^{-1}\,yr^{-1}}$ for the canonical and top-heavy IMF, 
    respectively.
 \end{enumerate}
These two points combined explain the different merger rates.

In calculating their rates, \citet{2018ApJ...856..140H} assumed that there exists a steady-state population of BH-BH binaries
in the central 0.1 pc, with a number of $N_\mathrm{steady}=200$ corresponding to a merger rate of
$2\,\mathrm{Gpc^{-1}\,yr^{-1}}$. This requires replenishment of the BH-BH binaries on a time-scale of 100 Myr,
which would be consistent with in-situ star-formation in a disk of massive O/B stars close to the MBH (e.g., \citealt{2009ApJ...697.1741B}).
In our rate calculation, we assumed a global influx of compact objects, which is more appropriate in our situation given that
we considered a binary population at larger distances from the MBH.

\citet{2016ApJ...828...77V} were the first to consider the combined effects of
VRR and LK evolution by means of hybrid $N$-body methods. Due to computational
limitations, the MBH masses were limited to up to
$10^4\,\msun$. \citet{2016ApJ...828...77V} found high merger fractions in their
simulations. This is consistent with our results, in the sense that we find
that the enhancement due to VRR is most significant for low MBH masses. More
quantitatively, \citet{2016ApJ...828...77V} found a merger fraction of
$\approx 0.25$ in their simulation with $M_\bullet=10^4\,\msun$ (10k
stars). This is very similar to our merger fraction of $\simeq 0.29$ for the
case $M_\bullet=10^4\,\msun$ and $\gamma=2$ (adding the semisecular and
non-semisecular fractions). This apparent agreement may be fortuitous, however,
given that we did not consider a number of potentially important effects (see
\S~\ref{sect:discussion:uncertainties}). In particular, DF is an important
process in the simulations of \citet{2016ApJ...828...77V}; prior to their
merger, most merging binaries sink inwards to the MBH\@. DF could therefore be
important for bringing a binary from a regime with $\adpar \gg 1$, where LK-VRR
coupling is unimportant, to a region in which $\adpar \sim 1$, where LK-VRR
evolution can drive the binary to merge.

We remark that \citet{2016ApJ...828...77V} extrapolated their results to higher
MBH masses similar to the MBH in the GC, and found high merger rates of
$\sim 100 \,\mathrm{Gpc^{-3}\,yr^{-1}}$. Our results show that the merger
fraction decreases rapidly with increasing MBH mass. Therefore, one should be
cautious when extrapolating rates from lower to higher MBH masses. In
particular, the rate of $100 \,\mathrm{Gpc^{-3}\,yr^{-1}}$ found by
\citet{2016ApJ...828...77V} may have been overestimated.

\section{Conclusions}
\label{sect:conclusions}
We considered the effect of torques induced by stars around an MBH (specifically, VRR),
 on the LK dynamics of binaries near MBHs. We
showed that VRR can increase the rates of strong interactions in binaries
orbiting the MBH by inducing high mutual inclinations between the inner orbit
of the binary and the outer orbit of the binary around the MBH\@. Consequently,
the binary can be driven to high eccentricities through the LK mechanism, even
if the initial mutual inclination between the inner and outer orbits is
small. These strong interactions include orbital energy loss due to GW
emission, implying that VRR can enhance the rates of compact mergers in
galactic nuclei, with implications for the detection of GWs. Our main
conclusions are listed below.

\medskip \noindent 1. VRR can enhance the efficiency of the LK mechanism if the
VRR and LK time-scales are comparable. We explored the associated parameter
space in \S~\ref{sect:time_scales}, and found that the `regions of interest',
i.e., the parameter space in which the VRR and LK time-scales are similar, is
largest for small MBH masses ($\sim 10^4\,\msun$), steep stellar density
profiles ($\gamma \approx 2$; $\rho_\star \propto r^{-\gamma}$), and wide inner
binary orbits ($a_\oin\gtrsim 10\,\au$). This can be understood qualitatively
by noting that the VRR time-scale increases relatively to the LK time-scale
with increasing MBH mass. Furthermore, for tight inner binary orbits
($a_\oin \lesssim 1\, \au$), PN precession in the inner binary orbit quenches
LK oscillations, regardless of VRR\@. Also, the regions of interest are
typically far away from the MBH (close to the radius of the sphere of
influence), where octupole-order terms are unimportant.

\medskip \noindent 2. Using a simplified model for VRR, we numerically
integrated the secular equations of motion for the three-body binary-MBH
dynamics coupled with the effects of VRR on the barycenter of the outer binary
(\S~\ref{sect:dyn}). We determined the eccentricity distributions as a function
of the `adiabatic parameter' $\adpar \equiv t_\mathrm{LK}/t_\mathrm{VRR}$. If
$\adpar\ll 1$, then the binary-MBH system is effectively decoupled from the
torques induced by the stellar background, and the canonical LK
dynamics apply (in particular, only high eccentricities can be induced if the
initial mutual inclination is close to $90^\circ$). If $\adpar \sim 1$, then
the dynamics are typically chaotic, and very high eccentricities can be
attained on time-scales of $\sim 10-100\,t_\mathrm{LK}$. If $\adpar \gg 1$,
then LK oscillations are suppressed, although there is diffusive evolution of
the eccentricity on long time-scales, on the order of
$\adpar \, t_\mathrm{LK}$.

\medskip \noindent 3. We carried out Monte Carlo calculations of
BH-BH binaries in galactic nuclei in \S~\ref{sect:pop_syn}, taking into account the
coupled LK and VRR evolution, PN corrections, and binary
evaporation. Consistent with the time-scale arguments of
\S~\ref{sect:time_scales}, we found that VRR is effective at enhancing the
rates of BH mergers for low MBH masses and steep density
profiles. For $M_\bullet=10^4\,\msun$ and a density slope of $\gamma = 2$, the
merger enhancement in terms of the delay-time distribution (\F~\ref{fig:DTD}) is
typically a factor of $\sim 10$, up to $\sim 60$ compared to the case of
canonical LK dynamics. In both cases, the merger rates decrease with time
approximately as $t^{-1}$. The merger fractions in our simulations with VRR
included range from less than a per cent for high MBH masses
($M_\bullet \gtrsim 10^6\,\msun$), to up to $\simeq 30\%$ (for
$M_\bullet=10^4\,\msun$ and $\gamma=2$). These fractions are a factor of a few
to $\sim 10$ times larger compared to the situation in which VRR is not taken
into account.

\medskip \noindent 4. Although the merger of the inner BH-BH binary is triggered by
high eccentricity, when reaching the LIGO band the inner orbit eccentricity is
highly damped due to GW emission. Some signature remains in the distribution of
$e_{\oin,\,\mathrm{LIGO}}$, although it is small compared to other dynamical formation 
channels: the median
$e_{\oin,\,\mathrm{LIGO}}$ in the simulations with VRR included is
$\sim 10^{-5}$, compared to $\sim 10^{-6}$ in the canonical LK case (see \F~\ref{fig:e_distributions}).
 The highest $e_{\oin,\,\mathrm{LIGO}}$ are attained in systems that also entered
the semisecular regime, in which the double orbit averaging approximation
formally breaks down.

\medskip \noindent 5. Large uncertainties are involved in the process of
converting merger fractions found in the Monte Carlo simulations to
absolute BH-BH merger rates. Nevertheless, we estimated the merger
rates for the GC and for galactic nuclei with different MBH masses, by making
two limiting assumptions on the formation rate of compact objects. With our
assumptions, we find that the GC merger rates are about two orders of magnitude
lower than the lower limit inferred by LIGO,
$12 \, \mathrm{Gpc^{-3}\,yr^{-1}}$ \citep{2017PhRvL.118v1101A}, if only LK
dynamics are included. With the inclusion of VRR, the rate enhancement is only
small, i.e., a factor of $\sim 1.6$, suggesting that LK-VRR coupling cannot
explain merger rates for galactic nuclei with masses comparable to the MBH in
the GC\@. The absolute rates and the enhancement by VRR generally
increase with decreasing MBH mass. Nonetheless, even in our most optimistic scenario, the baseline 
  BH-BH merger rate for other MBH masses is small, and the enhancement by LK-VRR coupling is not
  large enough to increase the rate to well above the LIGO/VIRGO lower limit. 
For $M_\bullet = 10^4\,\msun$, the rate barely reaches $12\,\mathrm{Gpc^{-3}\,yr^{-1}}$, 
 and is an order of magnitude lower than the upper LIGO limit, 
$213  \, \mathrm{Gpc^{-3}\,yr^{-1}}$ \citep{2017PhRvL.118v1101A}.

\section*{Acknowledgements}
We thank Hagai Perets, Bence Kocsis, Lisa Randall, Scott Tremaine, and Bao-Minh Hoang for 
stimulating discussions and comments on our manuscript. We also thank the
anonymous referee for a helpful and detailed report. 
In memoriam Yoshihide Kozai (April 1 1928 --- February 5 2018). 
ASH gratefully acknowledges support from the Institute for Advanced Study, the Peter
Svennilson Membership, and from NASA grant NNX14AM24G. 
BB acknowledges support from the Schmidt Fellowship.
CP acknowledges support from the Gruber Foundation Fellowship and Jeffrey L. Bishop Fellowship. 
FA acknowledges support from an E. Rutherford fellowship (ST/P00492X/1) 
from the Science and Technology Facilities Council.

\end{document}